\documentclass[epj]{svjour}
\usepackage{graphics}
\begin{document}
\title{Quantum Memories}
\subtitle{A Review based on the European Integrated Project
``Qubit Applications (QAP)''}
\author{Christoph Simon \inst{1,9} \and Mikael Afzelius \inst{1} \and J\"{u}rgen Appel
\inst{2} \and Antoine Boyer de la Giroday \inst{6} \and
Samuel J. Dewhurst \inst{6} Nicolas Gisin \inst{1} \and
Chengyong Hu \inst{7} \and Fedor Jelezko \inst{3} \and
Stefan Kr\"{o}ll \inst{4} \and J\"{o}rg Helge M\"{u}ller
\inst{2} \and Joshua Nunn \inst{5} \and Eugene Polzik
\inst{2} \and John Rarity \inst{7} \and Hugues de
Riedmatten \inst{1} \and Wenjamin Rosenfeld \inst{8} \and
Andrew J. Shields \inst{6} \and Niklas Sk\"old \inst{6}
\and R. Mark Stevenson \inst{6} \and Robert Thew \inst{1}
\and Ian Walmsley \inst{5} \and Markus Weber \inst{8} \and
Harald Weinfurter \inst{8} \and J\"{o}rg Wrachtrup \inst{3}
\and Robert J. Young \inst{6}}                     
%
%
\institute{Group of Applied Physics, University of Geneva,
CH-1211 Geneva, Switzerland \and Niels Bohr Institute,
Copenhagen University, Blegdamsvej 17, 2100 K{\o}benhavn
{\O}, Denmark \and 3. Physikalisches Institut, Universit\"at Stuttgart, Stuttgart, Germany \and Department of Physics,
Lund University, Box 118, SE-221 00 Lund, Sweden \and
Clarendon Laboratory, University of Oxford, Oxford OX1 3PU, United Kingdom \and Toshiba Research Europe Limited, 208 Cambridge
Science Park, Cambridge CB4 0GZ, United Kingdom \and
Electrical and Electronic Engineering, University of
Bristol, Bristol BS8 1UB, United Kingdom \and
Ludwig-Maximilians-Universit\"at M\"unchen, D-80799
M\"unchen, Germany and Max-Planck-Institut f\"ur
Quantenoptik, D-85748 Garching, Germany \and Institute for
Quantum Information Science and Department of Physics and
Astronomy, University of Calgary, Calgary T2N 1N4, Canada}
\date{Received: date / Revised version: date}
%
\abstract{We perform a review of various approaches to the
implementation of quantum memories, with an emphasis on activities within the quantum
memory sub-project of the EU Integrated Project ``Qubit
Applications''. We begin with a brief overview over
different applications for quantum memories and different
types of quantum memories. We discuss the most important
criteria for assessing quantum memory performance and the
most important physical requirements. Then we review the
different approaches represented in ``Qubit Applications''
in some detail. They include solid-state atomic ensembles,
NV centers, quantum dots, single atoms, atomic gases and
optical phonons in diamond. We compare the different
approaches using the discussed criteria. }
%

\maketitle

\section{Introduction}

Quantum memories are important elements for quantum
information processing applications such as quantum
networks \cite{KimbleNature}, quantum repeaters
\cite{Briegel,SangouardRMP} and linear optics quantum
computing \cite{KLM,KokRMP}. The field of quantum memories
has recently been very active. Recent reviews include Ref.
\cite{Lvovsky}, which gives a compact overview over
different memory protocols with atomic ensembles, Ref.
\cite{HammererRMP}, which deals with ensemble-based
memories with a detailed focus on atomic gases, and Ref.
\cite{TittelReview}, which deals with photon-echo based
quantum memories, in particular in solid-state atomic
ensembles, as well as Ref. \cite{LukinRMP}, which is
focused on quantum memories based on electromagnetically
induced transparency. Quantum memories were also discussed
in the Strategic Report on quantum information processing
and communication in Europe \cite{ZollerQIPC}. The high
activity of the field was also reflected in the European
Integrated Project {\it Qubit Applications (QAP)}, which
had a sub-project involving seven research groups devoted
to the development of quantum memories. As members of this
sub-project we made a sustained effort over the course of
the project to compare our various approaches. This
included the definition of appropriate criteria for such
comparisons. The present paper has grown out of this
effort. We hope that it will be useful for a wider
audience.

The approaches to quantum memories represented in the
project are quite diverse, including solid-state atomic
ensembles in rare-earth doped crystals, NV centers in
diamond, semiconductor quantum dots, single trapped atoms,
room-temperature and cold atomic gases, and optical phonons
in bulk diamond. It was therefore an interesting challenge
to develop a common language, in order to be able to
perform a meaningful comparison. We will begin this review
by discussing different applications for quantum memories
in section \ref{applications}, followed by a brief overview
of the different meanings that the term ``quantum memory''
can have in section \ref{types}. We will then review the
most important criteria for assessing the performance of
quantum memories in section \ref{criteria}. Section
\ref{requirements} discusses two physical requirements that
are important for good quantum memory performance in a
range of different approaches, namely good light-matter
coupling and good coherence. Section \ref{implementations},
which is divided in seven sub-sections, is devoted to an
overview of the different physical approaches that were
represented in {\it Qubit Applications}. In section
\ref{summary} we summarize our findings. This section
includes a comparative table that positions the different
approaches with respect to a number of relevant criteria.

\section{Applications of quantum memories}
\label{applications}

Quantum memories are important in a number of contexts,
including the implementation of single-photon sources,
quantum repeaters, loophole-free Bell inequality tests,
communication complexity protocols and precision
measurements.

\subsection{Deterministic Single Photon Sources}

One simple, but potentially very important application of a
quantum memory is the implementation of a single-photon
source. Given a non-deterministic source of photon pairs
such as parametric down-conversion, one can realize a
deterministic source of single photons by detecting one
photon of a pair (when it is emitted), while storing the
other one in a memory. Detection of the first photon
signals that the memory has been ``charged'' and can now be
used as a single-photon source. In order to implement a
close to ideal single-photon source the memory has to be
highly efficient. Deterministic single photon sources are
important ingredients for linear optics quantum computing
\cite{KLM,KokRMP}. They are also useful for certain quantum
repeater protocols \cite{SangouardRMP,SPS}.

\subsection{Quantum Repeaters}

Another area where quantum memories are absolutely
essential is for the implementation of long-distance
quantum communication via quantum repeaters. The direct
distribution of quantum states is limited by unavoidable
transmission losses in combination with the no-cloning
theorem. The quantum repeater approach \cite{Briegel}
overcomes this difficulty by combining quantum
teleportation and quantum state storage. The idea is to
divide a given long distance into shorter elementary links,
create and store entanglement independently for each link,
then extend it to the whole distance via entanglement
swapping (i.e. quantum teleportation of entanglement). This
approach is possible only if quantum memories are
available. Requirements for quantum memories in this
context include long coherence times, efficient interfacing
with photons (which serve as long-distance carriers of
quantum information) and the possibility of performing
entanglement swapping operations \cite{SangouardRMP}.
However, it is not absolutely necessary for the memories to be able
to both absorb and emit photons as we assumed in our
previous example for realizing a single-photon source.
There are different possible approaches, cf. the next
section.

\subsection{Loophole-free Bell test}

A more fundamental application for quantum memories is the
realization of a loophole-free test of Bell's inequality. A
loophole-free test requires both creating entanglement over
a long distance in order to enforce the locality condition
and detecting the entangled systems with high efficiency in
order to close the efficiency loophole. One attractive
approach towards achieving this goal is to create
entanglement between distant atoms or ions via the
detection of photons \cite{Saucke,SimonIrvine}. These atoms
or ions can be seen as quantum memories. Their interest in
the present context lies in the fact that they can be
detected with high (essentially unit) efficiency. At the
same time such a setup can also be seen as an elementary
link of a quantum repeater.

\subsection{Communication complexity and protocols requiring Local
  Operations and Classical Communication (LOCC)}
  Communication complexity, i.e. the number of qubits needed for
  achieving a certain communication task, has recently been shown to be
  tightly linked to the violation of Bell-type inequalities
  \cite{Buhrman:2010}. It may be useful to store quantum states
  transmitted by light in memories in order to achieve loophole free
  violations of this kind.  Communication protocols which require LOCC
  use memories to perform local operations and store the results while
  classical communication is going on. Examples of such protocols for
  continuous variables include iterative continuous variable
  entanglement distillation \cite{Browne2003}, continuous variable
  cluster state quantum computation \cite{menicucci:110501},
  communication/cryptography protocols involving several rounds
  \cite{Lamoureux2005}, and quantum illumination \cite{Lloyd:2008}.

\subsection{Precision measurements}
  A good quantum memory is capable of storing long-lived entanglement.
  As such, it can be used as a resource for entanglement enhanced sensing and
  precision metrology applications. A standard technique to enhance the
  weak coupling between light and matter is to
  use ensembles of many atoms. This approach does not only
  promise efficient quantum memories \cite{julsgaard04:_memory},
  \cite{DLCZ}, \cite{fleischhauer02:_quantum_memory_for_photons},
  but it can also increase the precision when detecting external
  disturbances on the state of such a system with quantum mechanically
  limited resolution. By preparing a collective entangled state of
  the atomic ensemble (e.g.  by storing a squeezed state of light in
  the memory\cite{appel:093602,honda} or by squeezing a collective atomic
  observable via a quantum-non-demolition (QND) measurement
  \cite{fernholz:073601}), the projection noise can be reduced.
  Due to the non-classical correlations between the atoms a
  measurement precision beyond the classical limit is possible. Already,
  spin squeezing on the Cesium clock transition has been
  demonstrated, which potentially can improve the precision of
  optical lattice clocks \cite{appel09:_spinsqueez}. Also, an ultra-sensitive
  RF-magnetometer employing entanglement in a room-temperature Cs
  vapour cell has been successfully
  developed \cite{wasilewski09:_magnet}. In NMR systems NOON
  states of 10 entangled spins have been employed to improve
  upon magnetic field sensing almost tenfold compared to
  standard measuring strategies \cite{jones01:_nmr}.

\section{Different types of quantum memories} \label{types}

It is useful to distinguish different ways of using quantum
memories and of characterizing their performance. This
includes the use of quantum memories for storing single
photons, for general states of light, and memories that are
charged through the emission of photons.

\subsection{Memories for single photons}
\label{singlephotonmemories}

For applications such as the implementation of a single
photon source, or quantum repeaters, one often knows that
the desired output state of the memory is a single-photon
state. For example, a quantum repeater protocol might
involve memories that are designed to absorb and reemit
photons that are emitted by other sources \cite{SPS,P2M3}
(e.g. photon-pair sources based on parametric
down-conversion, or more ideal single-photon or photon-pair
sources based on single atoms or quantum dots). In such a
situation, it is natural to verify the performance of the
memory using photon counting. The memory can then fail in
two distinct ways: it can fail to re-emit a photon at all,
or it can re-emit a photon whose quantum state has
imperfect overlap with the photon that was to be stored.
This motivates the distinction between efficiency
(probability to emit a photon) and fidelity (overlap of the
emitted photon with the original one). The latter is
sometimes more precisely denoted conditional fidelity,
since it is conditional on a photon having been emitted by
the memory. For quantum repeater applications, for example,
both efficiency and fidelity should typically be high
\cite{SangouardRMP}, however the requirements on the
efficiency tend to be somewhat less demanding than those on
the fidelity. It is interesting to note that for certain
ensemble-based implementations of quantum memories
decoherence can affect only the efficiency without
degrading the fidelity \cite{Staudt}. The memory
performance can also be characterized by quantum tomography
of the state of the emitted photon. The tomography
performed via a homodyne measurement is a deterministic
process which produces a full characterization of the
memory, from which efficiency and fidelity can be
extracted.

\subsection{Memories for general states of light}
One may also have the goal of storing general states of
light independently produced by a third party. Possible
applications range from linear optics quantum computing to
complex quantum communication protocols. Following the
storage of the state, the memory can either be
 measured in
some basis, or for other applications retrieved onto
another light pulse. Various types of states can be useful
here, such as coherent states, squeezed states,
Schr{\"o}\-dinger cat states, single photon, or Fock
states.  The characterization of the memory performance can
then be done either by quantum tomography of the atomic
state or by quantum tomography of the retrieved state of
light\cite{lobino:203601}.  Here again homodyne
measurements on light are used and thus one important
concept is the (unconditional) fidelity
\cite{PhysRevLett.94.150503}.

\subsection{Memories that emit photons and whose state can be
measured directly}

A different approach to quantum repeaters, but also to the
implementation of a loophole-free Bell test, cf. previous
section, involves memories (for example, single trapped
atoms) that emit photons which are entangled with the
memory, and where the atomic state is detected directly
\cite{Volz06,Moehring07,Rosenfeld07,Olmschenk09}. The
atomic state detection is typically a deterministic
process, so the distinction between efficiency and fidelity
does not apply in this case, the appropriate concept is
(unconditional) fidelity.

\subsection{Memories that emit photons and that are measured
via retrieval}

Even though they were not represented in QAP, we
would like to mention ensemble-based memories of the DLCZ
(Duan-Lukin-Cirac-Zoller) type \cite{DLCZ} where memory
excitations are created through the emission of a photon,
but where the final readout of the memory is done via
conversion of the memory excitation into a single photon.
Memory performance can then be characterized in analogy
with section \ref{singlephotonmemories}
\cite{SangouardRMP,HammererRMP}. \label{emissive}

\section{Criteria for assessing quantum memory performance}
\label{criteria}

We now discuss a number of evaluation criteria (figures of
merit) and their relevance for different applications and
implementations.

\subsection{Fidelity}

In general terms, fidelity is related to the overlap between the quantum state that is written into the memory and the state that is read out. It should be noted that the definition of fidelity that is typically used in the context of quantum memories is
different from the fidelity used in quantum information
theory, e.g. in the context of unambiguous state
discrimination, Uhlmann's theorem etc.
\cite{nielsenchuang}. For two pure states, the latter
corresponds to the modulus of their overlap, whereas the
former is the square of that modulus.

The precise operational meaning of the fidelity depends on the specific approach. For memories that store and re-emit single photons, the fidelity is defined as the overlap between the single-photon wave packet that was sent in and the one that is recovered from the memory. This is also denoted {\it conditional fidelity}, because it is conditional on the re-emission a photon. The probability to actually recover a photon is denoted {\it efficiency}, cf. the next subsection. For memories that are meant to store general states of light, the conditional fidelity is not an appropriate concept, and one has to consider {\it unconditional} fidelities.

Note that for some applications, the purity of
the output state (for a pure input) can be a better criterion than the fidelity. If the output state is related to the input state by a unitary transformation, this may do no harm. For example, a Bell
measurement involving two photon wave packets
\cite{braunsteinmann} that have both been stored in quantum
memories can be performed with high accuracy if both wave
packets have been subject to the same unitary
transformation (distortion). However, it seems that
impurity of the output state would be a problem for most
conceivable applications in quantum information. Such
impurity could have different forms, depending on the
implementation (e.g. timing jitter, depolarization).

It is clearly important to study which processes will limit the
fidelity (purity) of a given implementation. It is also
interesting to study which encodings of quantum information
are the most advantageous in a given physical system (are
there "decoherence-free subspaces"? \cite{lidar}).

\subsection{Efficiency}

We already introduced the concept of memory efficiency in the previous section. It is most clearly defined for the case of memories that are meant to store and emit single photons, where it is the probability to re-emit a photon that has been stored. In atomic ensembles the efficiency can in principle be close to one thanks to collective interference effects. For single atomic systems (including quantum dots and NV centers), the efficiency with which a single emitted photon can be recovered is typically small, but it can be enhanced through the use of optical cavities.
Efficiency is not a well-defined concept for memories that are meant to store general states of light, the use of (unconditional) fidelity is usually more appropriate.


It is interesting to note that, while high recall efficiency is
clearly desirable, it is not always necessary to be very
close to 100 \% for the memory to be useful, for example
for proof-of-principle demonstrations of quantum repeaters.
However, some applications, such as teleportation of an
unknown state, require that the storage of a light state in
the memory has unconditional high fidelity. In other words,
the combined fidelity times efficiency must be higher than
the classical limit. It is important to study the
efficiency required for different applications, see e.g.
Ref. \cite{SangouardRMP} for quantum repeater requirements.

\subsection{Storage time}

This is clearly very important for long-distance quantum
communication applications, where the communication time
between distant nodes imposes a lower bound on the required
storage time. Storage times have to be at least as long as
the average entanglement creation times, which are at least
in the second range for realistic repeater protocols and
relevant distances \cite{SangouardRMP}. A quantitative
study of the effect of storage time limitations was
recently performed in Ref. \cite{Lutkenhaus}. Long storage
times are less important for other applications such as
single-photon sources or loophole-free Bell tests.

\subsection{Bandwidth}

This will influence the practical usefulness of a given
memory for most applications, because it determines the
achievable repetition rates, and also the multiplexing
potential, cf. below. Bandwidth considerations can also be
important for matching sources (e.g. parametric
down-conversion) and memories. Of course the required
numerical value will depend on the desired application.

\subsection{Capacity to store multiple photons and dimensionality}

The capacity to store several modes is a natural capability
for certain ensemble implementations, which can allow a
multiplication of the repetition rate, e.g. in quantum
communication \cite{P2M3,Collins}. It is then of interest
to quantify the maximum number of photons (modes) that can
be stored. The criterion does not apply to single-atom
approaches in the same way. The ability to store multiple
spatial modes, i.e. to generate quantum holograms, which is
inherent to atomic ensembles is one exciting perspective
\cite{Polzikhologram}. For a single-mode memory, it can be
interesting to quantify its dimensionality, i.e. how many
excitations (photons) can be stored.

\subsection{Wavelength}

As mentioned above, for long-distance quantum communication
applications it is important that the wavelength of the
photons that propagate over long distances is within the
region of small absorption in optical fibers (unless one
considers free-space transmission, e.g. to satellites).
Depending on the protocol under consideration, this may
constrain the operating wavelength of the respective
quantum memory.

\section{Physical Requirements}
\label{requirements}

In this section we want to focus on the physical
requirements for implementing good memories, where "good"
is defined with respect to the figures of merit discussed
in the previous section. We will focus on figures of merit
that seem particularly important, and that are relevant for
all approaches, namely efficiency and/or fidelity, and
storage time. The described criteria are generally not
independent. For example, the longer the desired storage
time, the harder it may become to achieve high efficiency
and/or high fidelity.

We suggest that for any memory protocol there are at least
two requirements in order to achieve high
efficiency/fidelity and significant storage time, namely
good coupling between traveling and stationary excitations,
and good coherence.

\subsection{Light-matter coupling}

The notion of light-matter coupling is most easily defined
for memories that operate by absorbing and re-emitting
states of light (cases 2.1 and 2.2 above). In this case,
the notion of coupling is directly related to the
probability of absorbing the light. In the case of atomic
ensembles (whether in gaseous form or in solids), good
coupling is achieved by having high optical depth. If such
a memory is realized by a single quantum system inside a
high-finesse cavity, for example, then the finesse of the
cavity will play a role that is analogous to the optical
depth. For memories where the memory excitation is created
via the emission of a photon (cases 2.3 and 2.4), a
critical point is the probability to emit this photon into
a well-defined mode. It is at this point that the notion of
good coupling intervenes in such systems. For case 2.4, it
is also important in the readout process.

Optical depths that are sufficient for excellent memory
performance have already been achieved in atomic gas cells
(both hot and cold), and are in the process of being
achieved for rare-earth doped crystals, using multi-pass
configurations and/or particularly strongly absorbing
materials. High unconditional fidelity furthermore requires
good coherence and high detection efficiency (for the
homodyne detection), which have both been achieved in
experiments in Copenhagen (see section 6.5). For individual
systems such as trapped atoms and NV centers efficient
collection of the emitted photons could be achieved with
high-Q cavities. It is also worth noting that the
possibility of storing multiple modes (temporal, spatial,
directional) is being investigated in several
ensemble-based approaches. These points will be discussed
in detail in section 6.

\subsection{Coherence}

We also discuss decoherence. Its effects can be different
for different approaches. For example, for memory protocols
based on collective effects, such as controlled reversible
inhomogeneous broadening (CRIB), decoherence affects the
efficiency, but not the conditional fidelity. On the other
hand, for single atoms decoherence primarily affects the
fidelity, but not the readout or recall efficiency.
Nevertheless, decoherence is an important consideration for
all implementations. In most systems the coherence is
limited by fluctuating magnetic fields. These can be
externally applied fields (for trapped atoms) or fields
generated by the solid-state environment, in particular
nuclear spins in the host material (NV centers, quantum
dots, rare-earth ion ensembles). Coherence properties can
then be greatly improved by optimizing the choice of host
material (e.g. isotopically purified crystals that have
very low concentration of nuclear spins). The following
section develops the described topics in more detail for
the different experimental approaches within QAP.

\section{Implementations}
\label{implementations}

\subsection{Rare-earth ions in solids (Lund/Geneva)}

We now discuss the possible realizations of quantum
memories using ensembles of rare-earth ions in solids. The
optical 4$f$-4$f$ transitions in rare-earth ions are known
for their long optical coherence times (100$\mu$s-1ms)
\cite{Macfarlane2002,Macfarlane1987a} at cryogenic
temperatures and it is also possible to find spin
transitions with extremely long coherence time ($>$ 1
second) \cite{Fraval2004,Fraval2005,Longdell2005}. This
make them very suitable for transferring photonic quantum
states onto collective atomic coherences, which can be optical as well as spin. Doped into solids the optical transitions have
inhomogeneous broadening \cite{Macfarlane1990} much larger
than the homogenenous broadening given by the optical
coherence time. This causes inhomogeneous dephasing of the
induced optical coherence that needs to be compensated for.
We mention already that the inhomogeneous broadening offers
the possibility of a large-bandwith light-matter interface.

In order to control the inhomogeneous dephasing the
considered storage protocols are based on absorption and
re-emission of photons using photon-echo effects. It has
been shown, however, that traditional two-pulse photon
echoes are not suitable for the storage of quantum light
\cite{Ruggiero2009,ledingham-2009}. This is due to
intrinsic spontaneous-emission noise induced by the strong
$\pi$-pulse used for rephasing of dipole moments. We are
going to describe two modified photon echo protocols that
can in principle store single photons with efficiency and
fidelity up to unity. The first protocol is based on the
controlled and reversible inhomogeneous broadening (CRIB)
\cite{Moiseev2001,Nilsson2005,Kraus2006,Alexander2006,Sangouard2007,Alexander2007a,Hetet2008,Longdell2008,Lauritzen2009}
of a single absorption line (for a recent review see
\cite{TittelReview}). The second protocol is based on the
reversible absorption by a periodic structure of narrow
absorption peaks, a so called atomic frequency comb (AFC)
\cite{Afzelius2009a,Riedmatten2008,chaneliere-2009,Amari,Afzelius2009b}.
Potential areas of application for this type of quantum
memories are for long-distance quantum communication (quantum repeaters) and the realization of
quasi-deterministic single-photon sources.

\subsubsection{Controlled reversible inhomogeneous broadening}
We now describe the first scheme, controlled reversible
inhomogeneous broadening (CRIB) \cite{TittelReview}. This
involves broadening an initially narrow absorption line
using the linear Stark effect and an applied external
electric field gradient
\cite{Nilsson2005,Kraus2006,Sangouard2007,Longdell2008}.
The width of the broadened line should match the spectral
width of the light that is to be absorbed. Re-emission of
the light is triggered by changing the sign of the field,
which inverts the atomic transition frequencies around the
central frequency. The rephasing mechanism can be
understood by the following argument. Atoms at detuning
$\Delta$ with respect to the carrier frequency of the light
accumulates a phase $\exp(-i\Delta \tau)$ between
absorption ($t$=0) and the polarity inversion of the
electric field ($t=\tau$). By inverting the polarity, the
atoms invert their detunings $\Delta \rightarrow -\Delta$.
After a time $t=2\tau$ the atoms will have accumulated an
opposite phase factor $\exp(i\Delta \tau)$ which results in
a rephasing of all atomic dipoles.

The following formula can be derived for the memory
efficiency $\eta_{CRIB}$ \cite{Sangouard2007,Moiseev2004}:
\begin{eqnarray}
\eta_{CRIB}(t) = (1-e^{-\tilde{d}})^2
\mbox{sinc}^2(\gamma_0t),
\end{eqnarray}
where $\tilde{d}=d\gamma_0/\gamma$. Here $\gamma_0$ is the
width of the initial narrow line, $\gamma$ is the width of
the broadened line, $d$ is the optical depth of the initial
line, and $t$ is the storage time in the excited state.
Note that the initial absorption depth is decreased by the
broadening factor $\gamma/\gamma_0$. Thus the more
broadening that is added, the less efficient the storage
efficiency becomes. The storage of a single pulse (one
mode) can be chosen to be essentially equal to the pulse
duration (and thus inversely proportional to $\gamma$).
Once the pulse has been absorbed, the created atomic
excitation can be transferred to a lower-lying state
\cite{Nilsson2005,Kraus2006}, e.g. a state that lies in the
same hyperfine multiplet as the ground state. The storage
time in such a hyperfine state is limited only by
(hyperfine) decoherence, cf. below.

One can see that the total efficiency is the product of two
factors. The first one is related to the absorption of the
light. In fact, it is precisely the square of the
absorption probability in the broadened ensemble, since is
the optical depth of the broadened line. The square
intervenes because in the CRIB protocol, the re-emission
process is the exact time reversal of the absorption, so
the probability is the same for both processes. The second
factor (the squared sinc function) describes the dephasing
which is due to the fact that the initial narrow line has a
finite width $\gamma_0$. The exact form of the last factor
depend on the shape of initial line (which we here assume
to be a square function), it being related to its Fourier
transform.

In Ref. \cite{P2M3} an analysis of the performance of CRIB
was performed, particularly in terms of multimode storage.
It was found that an optical depth of about 30 is required
to achieve an efficiency of 0.9 for the storage of a single
pulse (one mode). If larger optical depths are available,
one can store multiple temporal modes (trains of pulses),
which can be very attractive for certain applications, e.g.
quantum repeaters. The number of modes that can be stored
with a certain fixed efficiency grows linearly with the
optical depth \cite{P2M3,Nunn2008}.

\subsubsection{Atomic Frequency Combs}
In order to fully exploit temporal multiplexing in quantum
repeater architectures, the memory should be able to store
many temporal modes with high efficiency \cite{P2M3}. For
EIT based quantum memories, this requires extremely high
and currently unrealistic value of optical depth
\cite{Afzelius2009a,Nunn2008}. The scaling is better for
CRIB based quantum memories but the required optical depth
are still very high (e.g. 3000 for 100 modes with 90$\%$
efficiency) \cite{P2M3,Nunn2008}. Recently, a new scheme
was proposed \cite{Afzelius2009a} for inhomogeneously
broadened materials, where the number of stored modes does
not depend on the initial optical depth $d$. The scheme is
based on atomic frequency combs (AFC).

\begin{figure}
\resizebox{0.9\columnwidth}{!}{\includegraphics{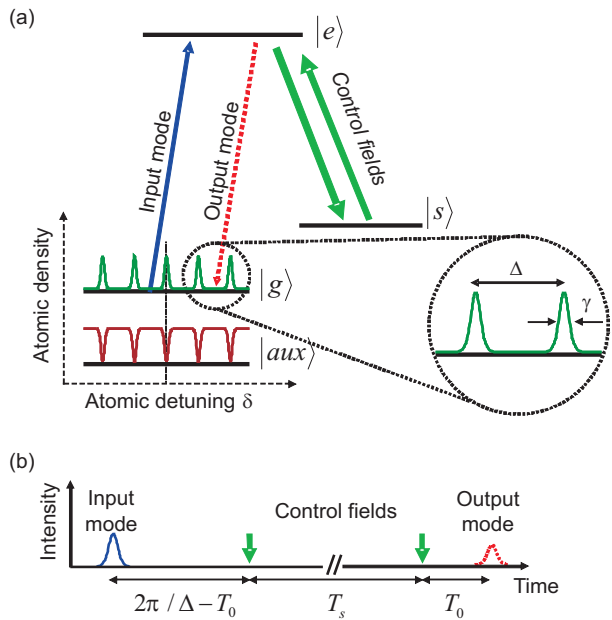}}\caption{The
AFC quantum memory scheme. Figure from Ref. \cite{Afzelius2009a}.} \label{AFC}
\end{figure}

The idea of AFC is to tailor the absorption profile of an
inhomogeneously broadened solid state atomic medium with a
series of periodic and narrow absorbing peaks of width
$\gamma_0$, height $d$ and separated by $\Delta$ (see Fig.
\ref{AFC}). The single photon to be stored is then
collectively absorbed by all the atoms in the comb, and the
state of the light is transferred to collective atomic
excitations at the optical transition. After absorption,
the atoms at different frequencies will dephase, but thanks
to the periodic structure of the absorption profile, a
rephasing occurs after a time $2\pi/\Delta$ which depends
on the comb spacing. When the atoms are all in phase again,
the light is re-emitted in the forward direction as a
result of a collective interference between all the
emitters.

In order to achieve longer storage times and on-demand
retrieval of the stored photons, the optical collective
excitation can be transferred to a long lived ground state
before the re-emission of the light. This transfer freezes
the evolution of the atomic dipoles, and the excitation is
stored as a collective spin wave for a programmable time.
The read out is achieved by transferring back the
excitation to the excited state where the rephasing of the
atomic dipoles takes place. If the two control fields are
applied in a counterpropagating way, the photon is
re-emitted backward. In that case, it has been shown that
the re-absorption of the light can be suppressed thanks to
a collective quantum interference.  In that configuration,
the theoretical storage and retrieval efficiency, assuming
that the decoherence in the long lived ground state is
negligible and a perfect transfer, is given by
\cite{Afzelius2009a}:
\begin{equation}
\eta_{AFC}\approx\left(1-e^{-\tilde{d}}\right
)^2e^{-\frac{7}{F^2}}
\end{equation}
where $\tilde{d} \approx d/F$ is the effective optical
depth and $F=\Delta/\gamma_0$ is the finesse of the AFC.
Note that this formula applied to a series of Gaussian
peaks. In Ref. \cite{chaneliere-2009} a formula for
Lorentzian line shapes is also given. In the equation above
we see that $\eta_{AFC}$ tends towards unity for
sufficiently large $d$ and $F$. Note that the
interpretation of the efficiency formula for AFC is
identical to the one for CRIB discussed above. The
difference lies in the origin of the effective absorption
depth $\tilde{d}$, which for CRIB is due to the applied
broadening while for AFC it is due to the removal of atoms
when creating the comb structure.

The number of temporal modes $N_m$ that can be stored in an
AFC quantum memory is proportional to the ratio between the
storage time in the excited state $2\pi/\Delta$ and the
duration of the stored photons, which is inversely
proportional to the total AFC bandwidth $\Gamma=N_p\Delta$,
where $N_p$ is the total number of peaks in the AFC. Hence,
we see that $N_m$ is proportional to $N_p$ and is
independent of the optical depth. See \cite{Afzelius2009a}
for a more detailed discussion on the multimode capacity.

\subsubsection{Crystal properties}
The optical depth of a given rare-earth doped crystal is
proportional to the dopant concentration (for low
concentrations) and to the transition dipole moment. Above
a certain dopant concentration, however, adding more ions
does not increase the optical depth any further
\cite{Koenz2003}, because ion-ion interactions lead to
increased inhomogeneous broadening without changing the
spectral density of absorbers at any given frequency.
Furthermore the optical coherence time may decrease for higher concentrations \cite{Bottger2006a}, equally due to ion-ion interactions. Of course, the optical depth can also be increased by using longer crystals, or, more significantly, by using
multi-pass configurations or optical cavities.

At low temperature ($T$=1-4 K) the optical and hyperfine
coherence times are limited by the spin-spin interaction of
the ions between themselves, and with spins in the host
crystal \cite{Macfarlane2002,Macfarlane1987a}. This has led
to the use of crystal hosts with low nuclear spin density,
most notably Y$_2$SiO$_5$. The commonly used ions
Praseodymium (606 nm), Europium (580 nm) and Thulium (793
nm) have quenched electronic spins, resulting in weak spin
interaction in general. For Erbium (1530 nm) and Neodymium
(880 nm) doped crystals, however, one has to take the (much
larger) electronic spins of Er and Nd into account, which
will interact both with nuclear spins in the host and among
themselves. Nevertheless, optical coherence times of a few
ms have been demonstrated for Er-doped \cite{Bottger2009}
and Eu-doped Y$_2$SiO$_5$ \cite{Koenz2003} (hundreds of
microseconds for Nd and Pr in the same host
\cite{Bottger2009}), while for Pr:Y$_2$SiO$_5$ hyperfine
spin coherence times as long as 1 second have been reported
\cite{Fraval2004,Longdell2005}.

In order to perform the spectral shaping of the
inhomogeneous absorption profile, which is necessary for
both CRIB and AFC, one uses optical pumping
\cite{Pryde2000,Nilsson2004,Rippe2005,Lauritzen2008}. The
optically pumped ions are stored in a ground state,
normally a spin level with a long population lifetime. Both
the CRIB and AFC protocols also use another spin-level for
storing the excitation on long time scales ($>$10 $\mu$s)
\cite{Moiseev2001,Nilsson2005,Kraus2006,Afzelius2009a,Afzelius2009b}.
In the preparation of the memory this level must be emptied
of population, also via optical pumping. Thus, the material
must have at least 3 spin levels for implementation of the
complete schemes. In materials with only 2 spin levels one
uses a level for population storage in the spectral shaping
process, and the light is only stored in the optically
excited state.

\subsubsection{Experimental state of the art}

The first proof of principle demonstration of the CRIB
scheme with bright coherent states was realized in a
Europium  doped solid \cite{Alexander2006}. The Europium
ions doped in the solid state matrix have an optical
transition at 580 nm, and a level structure with three
hyperfine states in the ground and excited states. The
authors used optical pumping techniques to create a narrow
absorption peak with a width of 25 kHz, within a 3 MHz wide
transparency window. The absorption of the peak was
approximately 40 $\%$. This peak was then broadened with a
gradient of electric field implemented with four electrodes
in a quadrupole configuration, thanks to the linear Stark
effect. An incoming optical pulse of 1 $\mu s$ was partly
absorbed by the broadened spectral feature. To read-out the
memory the polarity of the electric field was reversed
after a time $\tau$. After an additional delay $\tau$,
two-level Stark echoes were observed, with a decay time of
about 20 $\mu s$. In an another experiment, the same
authors stored and recalled a train of 4 pulses
\cite{Alexander2007a}. They also showed that the phase
information of the input pulses was preserved during the
storage. In these experiments, only a very small part of
the incident pulses were re-emitted in the Stark echo
(between $10^{-5}$ and $10^{-6}$). This low efficiency can
be partly explained by the small absorption of the
broadened peak (about $1\%$ ). In a more recent experiment,
the same group demonstrated a greatly improved storage and
retrieval efficiency of 15 $\%$ using a Praseodymium doped
crystal, which features an optical transition with larger
oscillator strength and consequently larger absorption
\cite{Hetet2008}. Using a longer crystal they could achieve
an efficiency as high as $45 \%$. Very recently, a CRIB
experiment has been demonstrated at the single photon
level, using an Erbium doped crystal absorbing at the
telecommunication wavelength of
1536 nm \cite{Lauritzen2009}.\\

Storage schemes combining off-resonant Raman interaction
and photon-echo rephasing mechanisms have been proposed
\cite{MoiseevTittel,HetetOL,LeGouetBerman}. The experiments
reported in \cite{HetetOL,Hosseini} use a Raman interaction
to transfer the optical input pulse onto a spin excitation.
The spin transition is artificially broadened by an
external magnetic field gradient, which allows rephasing of
the spin coherence using the CRIB scheme. These experiments
were realized in atomic vapours of Rb and in
\cite{Hosseini} an efficiency of 41\% was reached.
Moreover, it was shown that using this approach multiple
pulses could be stored and later retrieved in any time
order.

Phase preservation properties of photon echo techniques
have also been investigated for the storage of multiple
temporal modes using conventional photon echoes in Er doped
LiNbO3 \cite{Staudt2007a,Staudt}. This realizes a
(relatively low-efficiency) memory in the classical regime.
However, the results concerning the phase coherence
generalize to the case of CRIB memories. The phase
coherence of the memories was also studied performing an
interference experiment between two spatially separated
crystals \cite{Staudt}. The results showed very good
visibility (of order 90\%), which suggests that very high
fidelities should be achievable with CRIB and AFC based
memories in the same system. An interesting feature of
these results is that the decoherence only affects the
efficiency of the memory, but not its fidelity
\cite{Staudt} (as long as background fluorescence is
negligible). This is because atoms that are decohered
simply do not efficiently contribute to the collective
interference that produces the echo. The same principle
also applies to CRIB and AFC.

\begin{figure}
\resizebox{\columnwidth}{!}{\includegraphics{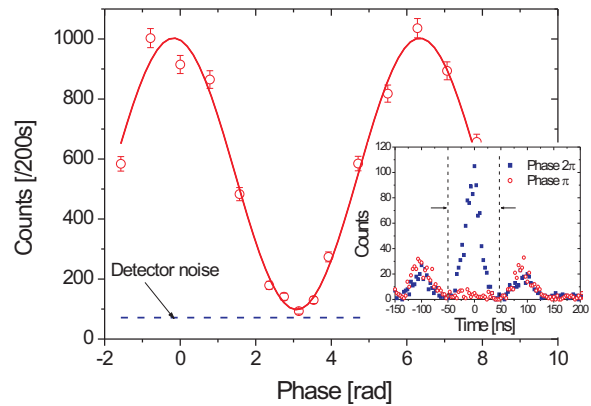}}
\caption{Phase preservation during the storage of a
time-bin qubit by an atomic frequency comb
\cite{Riedmatten2008}. Time-bin qubits with different
phases $\Phi$, are stored and analyzed using the
light-matter interface. The analysis is performed by
projecting the time-bin qubit on a fixed superposition
basis, which here is achieved by two partial readouts (see
text for details). The inset shows the histogram of arrival
times, where there is constructive interference for
$\Phi=2\pi$ and destructive interference for $\Phi=\pi$ in
the middle time bin. For this particular interference
fringe, a raw visibility of 82$\%$ and a visibility of
95$\%$ when subtracting detector dark counts are obtained. Figure from Ref. \cite{Riedmatten2008}.}
\label{phaseAFC}
\end{figure}

The first proof of principle demonstration of the AFC
protocol has been realized with a neodymium doped
crystal(Nd:YVO$_4$) \cite{Riedmatten2008}. This experiment
was the first demonstration of a solid light matter
interface at the single photon level. The authors
demonstrated the coherent and reversible mapping of weak
light fields with less than one photon per pulse on average
onto the ensemble of Neodymium ions. This experiment showed
that the quantum coherence of the incident weak light field
was almost perfectly conserved during the storage, as
demonstrated by performing an interference experiment with
a stored time-bin qubit (see Fig. \ref{phaseAFC}). Finally,
they also demonstrated experimentally that the interface
makes it possible to store light in multiple temporal modes
(4 modes). The storage and retrieval efficiency was low
(about 0.5 $\%$) in this experiment, mainly limited by the
imperfect preparation of the atomic frequency comb due to
the difficulty to perform good optical pumping in
Nd:YVO$_4$.
In a more recent experiment, \cite{chaneliere-2009}
demonstrated improved performance in terms of efficiency (9
$\%$) using a Thulium doped YAG crystal, also at the single
photon level. The efficiency could be further improved (to
17\%) by optimization of the creation of the AFC,
particularly by optimizing the shape of the teeth in the
AFC \cite{Bonarota}. The efficiency has been further
improved in Ref. \cite{Amari}, in the bright pulse regime,
reaching 34\% in a Pr-doped Y$_2$SiO$_5$ crystal. In the
weak pulse regime (average photon number $<1$), a storage
efficiency of 25\% was achieved \cite{Sabooni}. This
improvement was possible due to the ability to make narrow
(100-200 kHz) peaks with high absorption depth ($d\sim10$).
The multimode capacity has also been increased by almost
one order of magnitude compared to \cite{Riedmatten2008} in
a Nd-doped Y$_2$SiO$_5$ crystal, where up to 64 modes were
stored for 1.5 $\mu$s \cite{Usmani2009}.

In the AFC experiments cited above, only the first part of
the proposed AFC protocol was demonstrated  (i.e. the
coherent mapping onto collective excitation at the optical
transition and collective re-emission at a predetermined
time). Hence these experiments did not allow for on
demand-read out. Very recently, a proof of principle
demonstration of the full AFC protocol including the
transfer to a long lived ground state has been demonstrated
in a Praseodymium doped Y$_2$SiO$_5$ crystal
\cite{Afzelius2009b}. True on-demand storage up to 20$\mu$s
was achieved, limited by inhomogeneous spin dephasing. Spin
echo techniques should allow to extend the storage time up
to the spin coherence time (about 1 s) \cite{Longdell2005},
even at the single excitation regime.

\subsubsection{Performances}

We now evaluate the expected performances of quantum
storage in rare-earth doped solids with respect to the
parameters proposed in the introduction.

{\bf Fidelity:} Very high fidelities have been shown
possible since the dominant decoherence processes affect
the recall efficiency, but not the fidelity of the output
state. This is a virtue of ensemble-based quantum memories.
Experiments at the single photon level has shown
conditional fidelities up to 95\%.

{\bf Efficiency:} In theory efficiencies up to 100 \% are
possible with the CRIB and AFC protocols. In practice, the
maximal efficiency reached so far is 45$\%$ for the CRIB
protocol, and 34$\%$ for the AFC protocol, both in Pr-doped
Y$_2$SiO$_5$ crystals. The challenge is to design systems
with sufficiently high absorption. Another promising route
towards higher efficiencies is the use of multi-pass
arrangements or cavity enhanced light-matter interactions.

{\bf Storage time:} Rare-earth doped solids usually feature
optical transitions homogeneous linewidths of order of 1-10
kHz. Optical coherence times up to several ms have even
been observed in certain rare-earth doped materials (see
above). However, the storage time in the excited state is
mostly limited by the width of the absorption peak
$\gamma_0$. While the theoretical ultimate limit for the
width of absorption peak is given by the homogeneous
linewidth, in practice the narrowest peak realized so far
are or order of tens to hundreds of kHz. This leads to
storage times of tens of microseconds. Significantly longer
storage times can be achieved if the excitation is
transferred to a long lived ground state. In particular Pr
doped solids have extremely long spin coherence times. Even
longer spin coherence time are expected for Eu doped solids
\cite{Alexander2007}. Long storage times should thus be
possible.

{\bf Bandwidth:} Bandwidths up to several hundreds of MHz
are realistic using the atomic frequency comb protocol. The
challenge is thus to find systems with sufficiently high
initial absorption to achieve acceptable efficiencies. The
limitation in bandwidth will in general depend on the
spacing between hyperfine transitions, either in the
excited or ground state. Excitation of several levels will
cause quantum beat phenomena in the recall efficiency.

{\bf Multiple-photon and multiple-mode storage capacity:}
In principle, large numbers of photons could be stored in
the same memory. Trains of wave packets can be stored. The
AFC protocol is particularly promising for that aspect,
since the number of temporal modes that can be stored does
not depend on the available optical depth. In addition
there is potential for frequency multiplexing. Limits are
imposed by the storage time and the frequency bandwidth of
the material. In particular, the number of modes that can
be stored with the AFC protocol is proportional to the
number of peaks in the AFC.

{\bf Wavelength:} Rare-earth doped solids optical
transitions cover a wide range of wavelengths, from 580 nm
(Eu) to 1530 nm (Er). It should be noted that there is
considerable flexibility concerning the operating
wavelength of the photon memory in quantum repeater
protocols involving non-degenerate sources of entangled
photon pairs, since then photons at one wavelength can be
stored in the memory, while photons at another wavelength
can be sent into optical fibers. Some efficient quantum
repeaters proposals however require quantum storage at
telecommunication wavelengths. Erbium doped solids is the
only proposed system that meets the requirements to
implement a quantum memory at this wavelengths.

{\bf Read-out:} For the CRIB scheme, the read-out delay for
a single stored photon can be chosen at will by controlling
the electric field that causes the Stark shifts used in
CRIB. For the storage of multiple temporal modes, pulses
stored at different times will be re-emitted at different
times, thus allowing to distinguish between the modes. The
order of the pulse sequence is however reversed during the
storage. For the AFC scheme, the storage time in the
excited state is pre-determined by the comb structure. On
demand-read out can be achieved by a reversible transfer of
the stored excitations to ground state level. Contrary to
the CRIB scheme, the order of the pulse sequence is
preserved with the AFC scheme.

{\bf Complexity:} The proposed implementations uses
standard lasers and fiber-optic components. The lasers are
frequency stabilized to a reference for increased coherence
length, which is important when shaping the absorption
profile via optical pumping. The experiments also require a
cryostat, operating temperatures being in the region 1-4 K.

\subsection{NV Centers in Diamond (Stuttgart)}

\begin{figure}
\resizebox{0.5 \columnwidth}{!}{\includegraphics{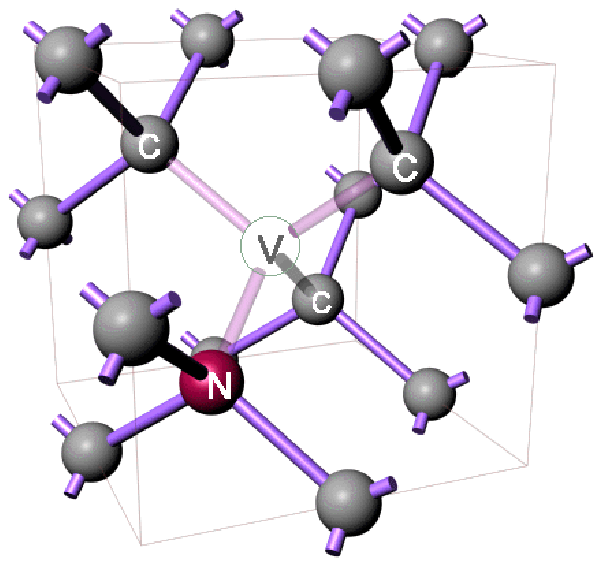}}
\resizebox{0.9 \columnwidth}{!}{\includegraphics{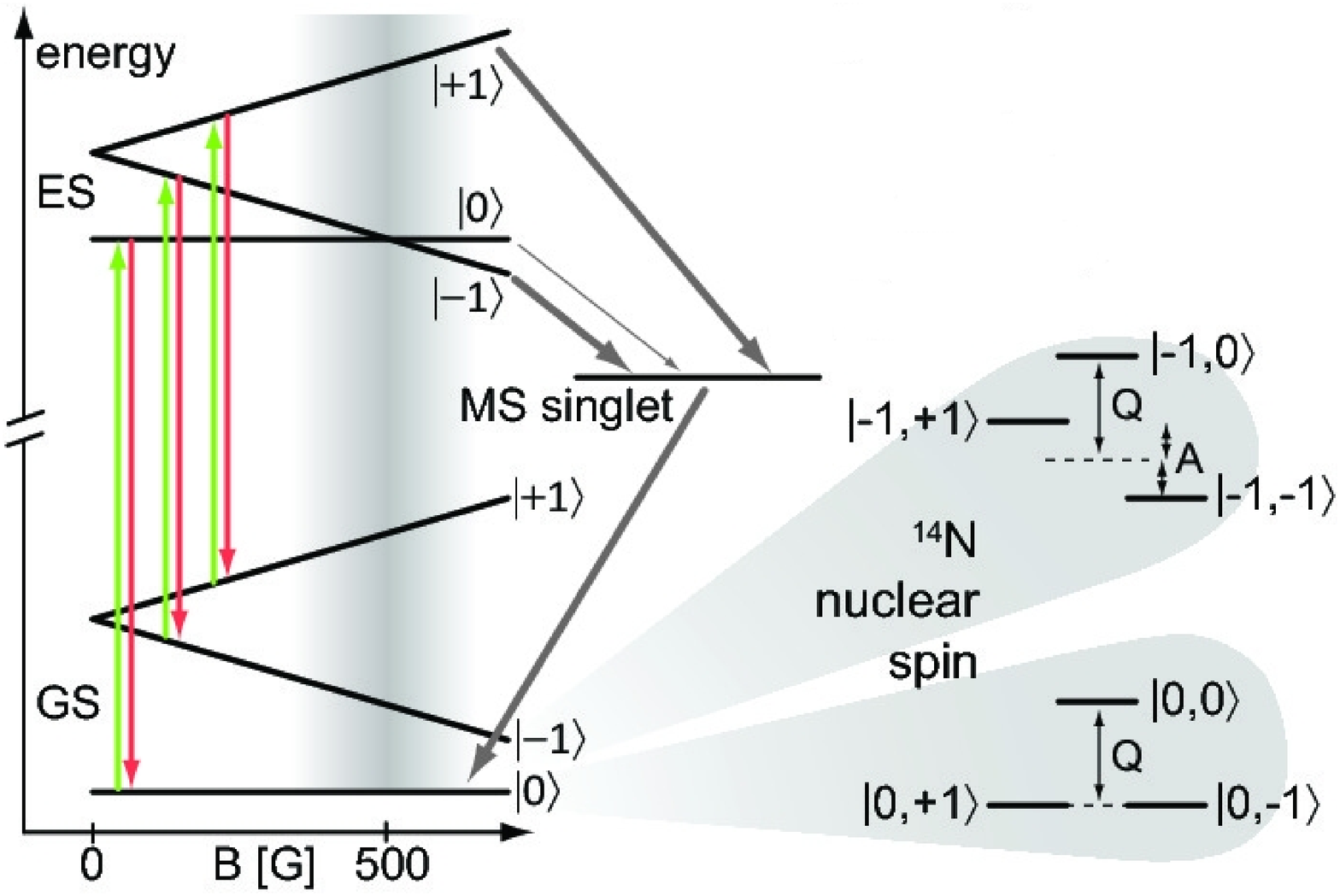}} \caption{In
(a) a schematic picture of the NV-center embedded in the
diamond lattice is shown. (b) sketches the NVs levels
scheme with its possible microwave transitions and
hyperfine structure as an inlay.} \label{fig1}
\end{figure}

Defect centers in diamond share many similarities with the
above mentioned rare earth system. Diamond defects
typically comprise single or few impurity atoms in a
diamond host with allowed optical transitions with
absorption wavelength raging from the UV to the infrared.
Those defects may be electron paramagnetic such that one
can use the superior relaxation and decoherence properties
of electron and nuclear spins for information storage,
retrieval and processing. In contrast to rare earth systems
up to now mainly defects have been studied in the context
of quantum information processing that do show strongly
allowed optical transitions (absorption cross section:
$10^{-17}$cm$^2$ at $T=300$K) such that single color
centers can be observed. The popularity diamond defects
have gained in quantum information over the past years is
based on two aspects: On the one hand color centers in
diamond usually are point defects deep within the bandgap
of diamond which results in narrow and stable room
temperature optical emission and excitation lines. In most
cases excitation and emission without lattice phonons being
involved is visible even at room temperature. On the other
hand the stiff and diamagnetic diamond lattice eventually
leads to very long electron and nuclear spin dephasing
times, which for example can reach ms for electron spins at
$T=300\,$K. The currently most studied defect center in
diamond is the nitrogen vacancy (NV) defect (see Fig.\
\ref{fig1}).

\begin{figure}
\resizebox{7cm}{!}{\includegraphics{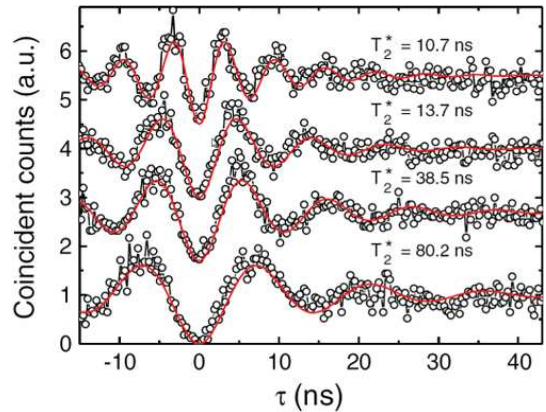}}
\caption{Second-order fluorescence intensity
autocorrelation function for the NV center at low
temperature under resonant excitation of $m_S = 0$ spin
state. The Rabi frequency of the excitation laser was
varied for different curves and equals to 0.44, 0.59, 0.67,
and 1.00 GHz (increase from bottom to top).} \label{fig2}
\end{figure}

Due to the fact that it is an electron paramagnetic system,
the NV does show a set of paramagnetic electronic and
hyperfine split ground and excited state levels. It thus
allows for the observation of photon-spin coupling used in
e.g.\ lambda-type optical transitions with the potential to
use CRISP and AFC type of techniques for storage. Long term
storage can be realized by electron and nuclear sub spin
levels with very long (ms-s) coherence times. The NV center
has a strongly allowed optical transition with a
corresponding optical excitation and emission linewidth of
around $10\,$MHz at low temperature. This enables strong
defect cavity coupling to enhance the spin photon interface
\cite{Park2006}. The project started with exploring the
fundamental applicability of the NV center as a quantum
storage element. Basically two lines of research were
explored namely an ensemble based approach for memory based
on EIT-like two-photon resonance. Secondly coupling of
single photon states to single NV centers which requires
efficient (strong) coupling to cavities is explored. An
important requirement prior to memory application is a
detailed understanding of the involved energy levels as
well as the excitation dynamics and photon properties of
the NV center. Fig.\ \ref{fig1} shows the energy levels of
the defect important for photon absorption and emission as
well as for electron spin physics. The optical excitation
path from the electron spin triplet ground state to the
excited state orbital doublet allows for electron spin
state conserving as well as spin flip transitions. If a
transition is of lambda type (spin flip transition) or not
can be determined by an external control parameter, e.g.\
transition wavelength or an external electric field. The
existence of lambda transitions has been proven in
electro-magnetic induced transparency type of experiments
\cite{Santori2006}. These experiments also demonstrate that
optical excitation can be used to generate long lived
ground state spin coherence, an important prerequisite for
quantum memory applications. A slightly different but key
observation connected to quantum repeater application as
well as light source of linear optical quantum information
processing is the fact that single NV defects do emit
transform limited photons \cite{Batalov2008} (see Fig.\
\ref{fig2}). To enhance spin-photon coupling single defect
centers have been coupled to plasmonic structures
\cite{Kolesov2009}.
Besides mere storage of photons via spin state coherence,
single defects in diamond also do possess the capability to
process the information stored. This particularly useful
property allows for e.g.\ entanglement purification in
quantum repeater applications. Information processing uses
$^{13}$C or nitrogen nuclei coupled to the electron spin of
the NV center. Coherence swapping between electron and
nuclear spins, entanglement among spins and basic quantum
gates have been demonstrated \cite{Jelezko2004}. It has
been estimated that 5 qbits are sufficient for a full
quantum repeater functionality \cite{Cappellaro2009}. This
proposed spin cluster comprises one electron spin for
spin-photon interaction and 4 nuclear spins for storage and
entanglement pumping. First experiments on clusters of such
size have been demonstrated in the meantime
\cite{Mizuochi2009}. Spin-photon entanglement has also recently been achieved.

\begin{description}
 \item[{\bf Fidelity:}]Fidelities for single or multiple photon storage in defect centers have not been determined up to now.
  However, subparts of the memory have been investigated towards their fidelity of state storage.
  This is particularly true for the spin memory part, i.e.\ swapping of coherences between electron and nuclear spin.
  Here fidelities as defined in the introductory part of this paper of up to $85\%$ have been achieved \cite{Dutt2007}.
 \item[{\bf Storage time:}]Storage time of the defect center memory is limited by the NV center $T_1$ or for arbitrary state storage $T_2$.
  In purified crystals $T_2$ for the electron spin has been measured to be ms.
  Ultimately information will be stored in nuclear spins with dephasing times on the order of tens of ms.
 \item[{\bf Bandwidth:}]Here the NV center can be used in a similar way than the rare earth systems in CRISP and AFC methods.
  A crucial part of these schemes is the possibility to create and reverse inhomogeneous broadening.
  In the case of the NV center this can be done by inverting a magnetic field or more importantly by the recently demonstrated Stark-shift frequency tuning \cite{Tamarat2006} allowing frequency multiplexing.
 \item[{\bf Wavelength:}]The NV center emission wavelength is centered around $700\,$nm and hence not in the telecommunication range.
  Nevertheless free space quantum communication was demonstrated with NV defects ($637\,$nm) \cite{Alleaume2004}.
  Emission at longer wavelength is achieved with different defect centers.
  Other defects in diamond like NE8 (emission wavelength $800\,$nm) might be more suitable for fiber-based systems \cite{Gaebel2004}.
  In addition to this defect, 90\% of oscillator strength is concentrated in zero-phonon transition of NE8 defect (compared with 10\% for NV defect).
  Currently however, little is known about the spin states of this defect and their coherence properties.
 \item[{\bf Complexity:}]
  For storage of single photons in defect center cavity systems first steps have been taken.
  As mentioned in the introductory part strong coupling between single defect in diamond and the whispering gallery mode of silica micro resonator was shown.
  The development of other cavity system is in its beginnings.
  However progress was reported on NV center photonic band gap cavity coupling as well as NV center-plasmon coupling.
  Owing to the low electron phonon coupling diamond-based photonic memory can be operated at relatively high temperatures (up to $20\,$K).
\end{description}

\subsection{Semiconductor nanotechnology (Toshiba/Bristol)}

Semiconductor quantum dots could potentially be used to produce fast and efficient quantum memories to be utilised in future quantum networks or as quantum repeaters. Semiconductor based systems are compact, robust and easy to integrate with existing technology. The energy structure of the quantum dots can be engineered by manipulating the material, size, and strain. Furthermore, the well established techniques to produce high-Q cavities in semiconductor material makes it possible to achieve a high efficiency \cite{Shields07}.

\begin{figure}
\resizebox{7.45cm}{!}{\includegraphics{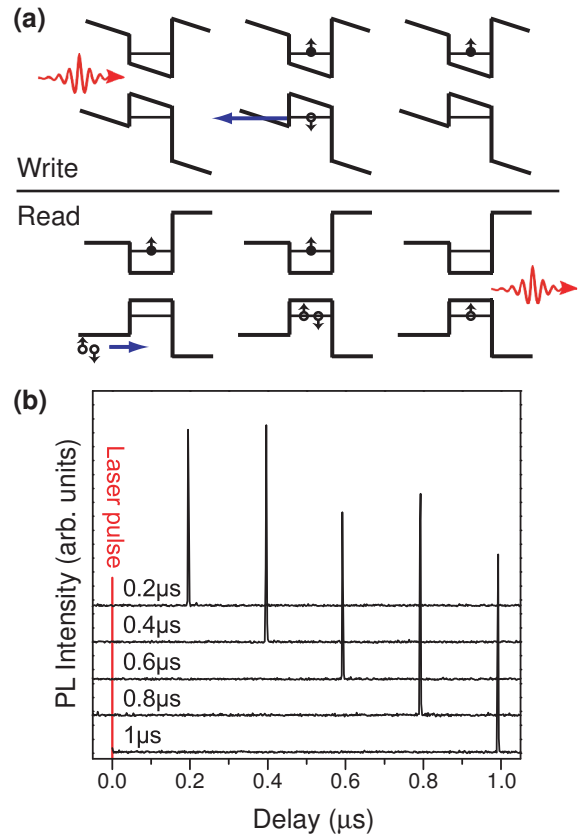}}
\caption{(a) Schematic diagram of the quantum dot band structure and the memory scheme for pure spin storage. Write mode: A circularly polarised photon is absorbed and creates an exciton, the device is biased to remove holes (open circles), the polarisation is stored in the spin-state of the confined electron (filled circles). Read mode: The device is biased to return a pair of holes, radiative recombination results in a photon with a polarisation dictated by the stored electron's spin state. (b) Time-resolved emission from the quantum dot for readout delay times between 0.2 and 1 $\mu$s (as labeled).}
\label{TREL1}
\end{figure}

In our approach we are using a single InAs quantum dot embedded in GaAs to store and deterministically re-emit a single-photon state. The operation of the device as described in \cite{Young07} is illustrated in Fig. ~\ref{TREL1} (a). The quantum dot is placed in the intrinsic region of a diode structure and biased so that the heavy-hole tunnelling rate is higher than the radiative recombination rate. On the negative side of the diode an AlGaAs barrier prevents the electron from leaving the dot. The polarisation of an incoming circularly polarised photon is transcribed to a pure spin state of an exciton confined in the quantum dot. The hole immediately tunnels out while the electron remains, thus storing the spin state. At a later time a pair of heavy-holes is returned by applying an ac voltage pulse. The positively charged exciton that is formed subsequently recombines, recreating a photon with a polarisation defined by the spin-state of the stored electron. Fig. ~\ref{TREL1} (b) shows the delayed emission for storage times up to 1 $\mu$s. The storage time is limited by the voltage pulse pattern generator and not the device, no decrease (other than statistical fluctuations) in the readout intensity can be observed on this time scale.

\begin{figure}
\resizebox{6.32 cm}{!}{\includegraphics{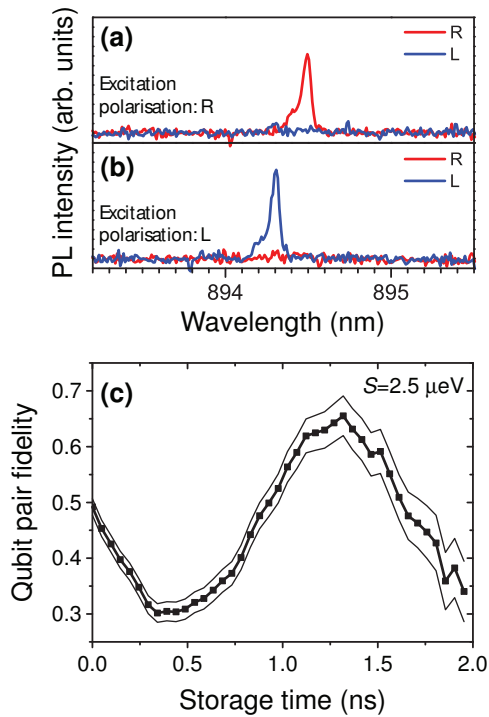}}
\caption{(a) and (b) The spectrum of the delayed emission after 1 $\mu$s storage of a right- and left-hand circularly polarised photon respectively. The spectrum is measured in the in the right- (R) and left-hand (L) polarisation basis. (c) Coherent time-evolution of the fidelity, $f^+$, for the exciton-photon state with the symmetric Bell state. Bands denote measurement errors dominated by Poissonian counting noise.}
\label{TREL2}
\end{figure}

{\bf Fidelity:} The fidelity for the pure spin storage described above was 80 $\pm$ 10 \% and constant within the delay times. This is consistent with the spin relaxation time for similar spin memories which has been shown to be in the ms range \cite{Kroutvar04}. Fig. ~\ref{TREL2} (a) shows the spectrum of the delayed emission after 1 $\mu$s storage of a right-hand circularly polarised photon. The spectrum is measured both in the right- (R) and left-hand (L) polarisation basis to illustrate the high fidelity of the readout. Fig. ~\ref{TREL2} (b) shows the same measurement for storage of a left-hand circularly polarised photon.

 In order to store a superposition state both the electron and hole must be retained (as their spin states are entangled), this can be done in separate reservoirs. In this case the fidelity is limited by the hyperfine interaction between the stored electron and the nuclei of the atoms which form the quantum dot (and those in the material surrounding the dot, depending on the extent of the electron's wavefunction). For the hole, which has a p-like orbital with almost no overlap with the nuclei, the hyperfine coupling is very small. The nuclear spins give rise to a randomly fluctuating magnetic field which causes the electron spin to pick up a random phase and thus the electron spin-coherence decays with time. For III-V semiconductor material (such as the InAs-GaAs system discussed here) the decoherence time can be expected to be of the order of 10 ns.

To demonstrate coherent storage we have created an entangled exciton-photon state in a quantum dot, and observed the coherent evolution of the spin-superposed state in a dot for up to 2 ns, with a single qubit fidelity of up to 94 \% \cite{Stevenson08}. The entangled exciton-photon state was prepared through the radiative decay of a biexciton in a single quantum dot. Polarisation selection rules dictate that radiative decay will project the system into the entangled exciton-photon state $\Psi \propto (|H_{XX}X_H\rangle+e^{iS\tau/\hbar}|V_{XX}X_V\rangle)$   where $H_{XX}$ and $V_{XX}$ are the vertically and horizontally polarised photon states and $X_H$ and $X_V$ are the exciton spin states coupling to the horizontally and vertically polarised photons respectively. $S$ is the spin splitting of the stored exciton state. Fig. ~\ref{TREL2} (c) shows the measured fidelity, $f^+$, of the exciton-photon state to the symmetric Bell state $\Psi^+=(|H_{XX}X_H\rangle+|V_{XX}X_V\rangle)/\sqrt{2}$, as a function of the storage time. Strong oscillations are observed as a function of the storage time with a period of $h/S$. The increasing phase acquired by the stationary qubit causes the final entangled photon pair state to periodically overlap with the symmetric Bell state. No loss of coherence is resolved for the longest recorded storage time of 2 ns which is limited by the radiative lifetime of the exciton. We stress that when the qubit pair fidelity is low, entanglement still exists in the system but with high fidelity to other Bell states.

Although the spin dephasing time is in the 10 ns range for III-V semiconductors, where the naturally occurring isotopes of the constituent materials all have non-zero nuclear spin \cite{Krebs08}, the nuclear field fluctuates on a much slower time scale, on the order of 100 $\mu$s. This means that a spin echo technique could be employed to extend the coherence time to this range \cite{Hanson07}. Another alternative is to polarise the nuclear spins by optical pumping to keep the nuclear field fixed in time. Thereby any uncertainty in the phase evolution of the electron spin would be removed \cite{Khaetskii02,Coish06}. It should also be mentioned that other material systems exist where most of the nuclei are zero spin isotopes. An example of such a material system is CdSe QDs embedded in ZnSe \cite{Akimov06}.

{\bf Efficiency:} For this memory scheme the input and output coupling efficiency for the photon to dot-confined exciton are similar and they predominantly determine the efficiency of the memory as a whole. Strong coupling between optical and exciton modes has been demonstrated by integrating quantum dots into optical cavities. To achieve strong coupling, the exciton-photon coupling parameter $g$ must meet the criteria: $g^2>(\gamma_c - \gamma_x)^2/16$ \cite{Reithmaier04},  where $\gamma_c$ and  $\gamma_x$ are the full width at half maxima of the cavity and exciton modes respectively; thus to reach the strong coupling limit it is necessary to fabricate cavities with very high quality factors (Q) to make the width of the cavity mode comparable to the exciton's mode. Pillar microcavities formed by etching through distributed Bragg reflectors, two-dimensional photonic crystals and microdisks have all been used to show strong exciton-photon coupling \cite{Reithmaier04,Hennessy07,Peter05}. Such systems should be capable of providing near deterministic single photon absorption and emission which is required for a useful quantum memory. Our best output coupling efficiency for microcavities is 20\%, a similar input coupling efficiency should in principle be achievable. To our knowledge the best quantum efficiency achieved for exciton-cavity mode coupling between excitons in quantum dots and a microcavity is 97 \% \cite{Press07}. We have been working to extend strong coupling schemes to show the possibility of a deterministic photon-spin quantum interface using charged dots \cite{HuPRB}. A singly charged quantum dot, when spin polarised will interact differently to different circular polisation states of light due to Pauli blocking. When this occurs in a strongly coupled QD-cavity system the spin state of the charged dot induces giant circular birefringence and giant optical Faraday rotation. This enables non-demolition measurement of single spins, entanglement between a spin and a single photon and the development of various photon-spin, photon-photon and spin-spin entangling gates. As a result this could provide an efficient read-in, read-out quantum memory and quantum relay. As yet no experimental demonstration has been reported.

{\bf Storage time:} The limiting factor here will be the decoherence time of the electron spin stored in the quantum dot (which could be $>100 \mu$s).

{\bf Bandwidth:} The bandwidth will be limited by the speed at which the gates controlling the device can be operated, hole tunnelling times, and the exciton's radiative lifetime. The latter is usually around 1 ns for an isolated quantum dot but can be greatly reduced for a quantum dot in a cavity by the Purcell effect. Operation at 1 GHz has been demonstrated for single photon emission from quantum dots in diodes \cite{Bennett05}.

{\bf Multiple-photon and multiple-mode storage capacity:} Frequency multiplexing is possible with multiple dots of different size. Multiple photon storage in the single quantum dot scheme is not likely.

{\bf Wavelength:} Initial experiments exploit the benefits of silicon detectors and work in the near-IR,  around 900 nm. We have developed some quantum dot based devices at telecom wavelengths \cite{Ward05}. In principle the emission wavelength can be tuned arbitrarily by using appropriate materials and manipulating the quantum dot size and strain.

{\bf Complexity:} Production of the quantum memories only involves standard semiconductor fabrication techniques. The operation of the quantum memory however requires liquid helium temperatures. As the temperature is increased confinement of the electron in the dot reduces as thermal activation becomes possible. This limits operation to $<$50 K for typical InAs quantum dots which are optically active at 900 nm. Confinement can be increased to allow higher temperature operation by altering the growth of the dots or changing the dot and/or barrier materials.

In summary, we have developed a spin memory based on semiconductor quantum dots. This was done by creating devices which provide preferential tunnelling of holes to allow a spin encoded exciton to be decomposed and the pure spin state stored by the electron confined in the dot. With this device we have demonstrated storage of pure spin states for 1 $\mu$s with a fidelity of 80 \%. Furthermore, we have demonstrated that the coherence of a photon entangled with a stored exciton does not decay within the exciton radiative lifetime. We also present possible routes to improve a semiconductor quantum memory.

\subsection{Single optically trapped $^{87}$Rb atoms (LMU
Munich)}

{\bf Brief description of approach and potential
applications:} This approach aims at the generation of
entanglement between remote atomic quantum memories useful
for future applications in long-distance quantum
communication like quantum networks or the quantum repeater
\cite{Briegel}. Such applications can be realized on basis
of our recently demonstrated entanglement between a single
optically trapped $^{87}$Rb atom and a single emitted
photon \cite{Volz06,Rosenfeld08} by applying robust
probabilistic quantum gates like entanglement swapping
\cite{Zukowski93}, based on the quantum interference of
photon pairs on a beam splitter
\cite{Moehring07,Matsukevich08,Maunz09}. The memory qubit
is stored in the Zeeman sub-levels $m_F = \pm 1$ of the
$5^2S_{1/2}, F=1$ hyperfine ground level of a single
$^{87}$Rb atom (localized in an optical dipole trap
\cite{Weber06}), whereas the photonic communication qubit
is encoded in the polarization state of a single photon.
Transfer of quantum information from the photon onto the
quantum memory can be realized for example by quantum
teleportation or similarly by remote state preparation
\cite{Rosenfeld07}. Efficient readout with high fidelity
can be performed in two ways: The memory qubit can be
determined by a destructive spin measurement using a
fluorescence-based shelving technique \cite{Volz06} or by
state-selective laser ionization and subsequent detection
of the charged ionization fragments. Both techniques are
investigated by our group \cite{Henkel09}. Alternatively,
stored quantum information can be coherently transferred
back onto a photonic qubit by optically exciting the atom
on a suitable dipole transition, leading to the emission of
a single photon in the same spin state as the atom. The
latter technique was recently demonstrated for the first
time by Wilk {\it et al.}, where a single $^{87}$Rb atom
emits a single photon into the well defined spatial mode of
an high-Q optical cavity \cite{Wilk07}.

\begin{figure}
\resizebox{0.95\columnwidth}{!}{\includegraphics{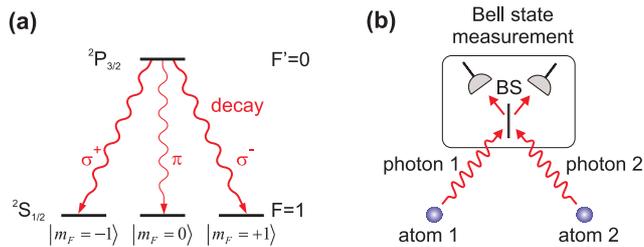}}
\caption{{\bf (a)} Preparation of atom-photon entanglement
in $^{87}$Rb via spontaneous Raman scattering. The atoms
are excited first with short optical laser pulses to the
$5^2P_{3/2}, F'=0$ hyperfine level. In the following
spontaneous emission process they decay to the ground
states with the magnetic quantum numbers $m_F=-1, 0,$ or
$+1$ by emitting a $\sigma^+, \pi,$ or $\sigma^-$-
polarized photon, respectively. Provided these decay
channels are indistinguishable in all other degrees of
freedom and the $\pi$-decay is filtered out spatially a
maximally entangled state is formed between the photonic
qubit $\{|\sigma^+\rangle, |\sigma^-\rangle\}$ and the
atomic spin-qubit $\{|m_F=-1\rangle, |m_F=1\rangle \}$.
{\bf (b)} Preparation of a pair of entangled atoms from two
entangled atom-photon pairs via entanglement swapping. The
photons interfere on a $50:50$ beam-splitter (BS). A
coincident detection of the photons in both output ports
projects the atoms onto an entangled state.}
\label{fig:entswap}
\end{figure}

For future applications of long-distance quantum
communication like quantum networks, the quantum repeater
\cite{Briegel}, or quantum teleportation between distant
matter qubits it is mandatory to achieve entanglement
between separated quantum processors. In order to entangle
two $^{87}$Rb atoms at remote locations, we intend to apply
the entanglement swapping protocol \cite{Zukowski93}. For
this purpose, both atoms are entangled in a first step via
spontaneous Raman scattering with a single photon
\cite{Volz06} (see Fig. \ref{fig:entswap} (a)). Each of the
two photons is collected via a large numerical aperture
objective, coupled into a single mode optical fiber, and
guided to an intermediate location. There the photons are
overlapped on a 50:50 beam-splitter (BS). In case the
photons are detected in coincidence in both output ports of
the BS the remaining atom-atom pair is projected onto the
maximally entangled spin-singlet state (see Fig.
\ref{fig:entswap} (b)). First successful experimental steps
into this direction -- namely the demonstration of
entanglement \cite{Moehring07,Matsukevich08} and quantum
teleportation \cite{Olmschenk09} between single-ion quantum
bits at a distance of 1 m -- were reported recently by
Chris Monroe's group.

In this contribution we focus on figures of merit which are
particularly important for the generation of entangled
$^{87}$Rb atoms connected via optical fiber links of
several 100 m length.

{\bf Efficiency:} The overall success probability of the
entanglement swapping protocol is given by
\begin{equation}
P_{At-At}= \frac{1}{4} (p \eta)^2 e^{-\alpha (L_1+L_2)},
\label{eqn:LMU1}
\end{equation}
where $p$ denotes the overall collection and coupling
efficiency into the optical fiber, $\eta$ the quantum
efficiency of the single photon detectors, and $e^{-\alpha
(L_1+L_2)}$ the photon attenuation in the optical fiber
links with communication distances $L_1$ and $L_2$ and
absorption $\alpha$ . The factor 4 accounts for the fact
that only one out of four photonic Bell-states is detected.
Typically $p\eta\approx 10^{-3}$, provided the photon
attenuation is neglected. In practice we expect to achieve
an overall success probability of $2 \times 10^{-7}$ for
the entanglement swapping process \cite{Rosenfeld09a} via a
300 m optical fiber link. This value could be enhanced by
the use of improved detection optics
\cite{Kurtsiefer08,Leuchs09} or optical cavities
\cite{Wilk07}, thereby allowing entanglement swapping at a
higher rate.

{\bf Fidelity:} When talking about fidelities (accuracies)
of this scheme we distinguish in general three cases. (i)
The fidelity $F_{At-Ph}$ we are able to generate an
entangled atom-photon state, (ii) the accuracy $a_{det}$ to
read out the atomic spin-state via a projective
measurement, and (iii) the fidelity $F_{At-At}$ to generate
a pair of entangled atoms at remote locations via
entanglement swapping. In theory atom-photon entanglement
fidelities $F_{At-Ph}$ of up to 100$\%$ are possible.
However in practice due to errors in the preparation of the
excited atomic state $5^2P_{3/2},F'=0$ the best achievable
fidelity is limited to $F_{At-Ph}=99.6 \%$. To estimate now
the expected atom-atom entanglement fidelity $F_{At-At}$
after one photon has traveled e.g. an optical fiber link of
300 m length, one has to account for additional errors
which are mainly due to polarization imperfections in the
optical fiber \cite{Rosenfeld08}, dephasing of the
entangled atom-photon state (caused by fluctuating magnetic
fields \cite{Rosenfeld08}), and spatio-temporal mismatch in
the two-photon interference. In addition, dark counts in
the single photon detectors of the Bell-state analyzer will
further reduce the achievable fidelity. Applying a careful
error analysis we estimate a fidelity of $F_{At-At}=94.4\%$
to generate an entangled pair of atoms \cite{Rosenfeld09a}.
Here we emphasize that this value yet does not include the
limited accuracy $a_{det}=95\%$ to distinguish orthogonal
atomic spin states via fluorescence-based shelving.
Including this additional experimental error we finally
estimate an observable entanglement fidelity of
$F^*_{At-At}=81\%$ \cite{Rosenfeld09a}.

{\bf Storage time:} If we scale this scenario to larger
distances (several km), dephasing of the atomic spin due to
fluctuating magnetic fields -- respectively the resulting
dephasing of the entangled atom-photon interface -- will
reduce the reachable atom-atom entanglement fidelity
$F_{At-At}$. By implementing an active magnetic field
stabilization and without application of any magnetic
guiding field we demonstrated transverse and longitudinal
spin-dephasing times of $T^*_2=75..150$ $\mu$s and $T_1\ge
0.5$ ms, respectively, limited mainly by the residual
state-dependence of the optical trapping potential and the
thermal energy of the trapped atom \cite{Rosenfeld08}. This
value is in good agreement with state-of-the-art coherence
times of Zeeman qubits stored in single trapped ions. In
order to achieve storage times up to several seconds one
could use e.g. magnetic-field-independent hyperfine qubits
and encode the photonic qubit in the frequency of a single
photon \cite{Moehring07}, thereby enabling e.g. quantum
teleportation between matter qubits \cite{Olmschenk09} at
distances of several km. Nevertheless, with the already
demonstrated coherence time of $T^*_2=75..150$ $\mu$s for
our Zeeman qubit an entanglement swapping experiment
between two atoms connected via 300 m optical fiber can be
performed in the near future.

{\bf Bandwidth:} The Fourier-limited bandwidth of our
single atom quantum memory is determined by the natural
linewidth (6 MHz) of the dipole transition $5^2S_{1/2}
\rightarrow 5^2P_{3/2}$ in $^{87}$Rb. This value assures a
rather uncritical implementation of the entanglement
swapping protocol since the interfering photon wave packets
have an overall spatial extension of several meter. In
addition the temporal shape of the photons could be tuned
e.g. via controlled variation of the spontaneous Raman
scattering process \cite{Eschner09} while maintaining the
possibility to generate a highly entangled atom-photon
interface.

{\bf Wavelength:} The wavelength $\lambda$ of the photonic
qubit is determined by the dipole transition $5^2S_{1/2}
\rightarrow 5^2P_{3/2}$ in $^{87}$Rb. At $\lambda=780$ nm
the damping in optical fibers is $\approx 5..6$ dB/km. In
combination with our single photon collection efficiency of
$p=4..5\times 10^{-3}$ our current experimental
implementation (two single atom traps) already allows the
distribution of quantum information and entanglement
between nodes of a primitive quantum network at a distance
of 1 km.

\subsection{Room temperature alkali gases with long spin polarization
  life time (Copenhagen)} \label{sec:room-temp-alkali}

  \paragraph{Brief description of approach and potential applications:}
  Room temperature atomic gases allow for the use of various
  approaches towards a quantum memory, such as Raman transitions
  \cite{Kozhekin} and electromagnetically induced
  transparency \cite{fleischhauer02:_quantum_memory_for_photons}. The
  approach we have been pursuing is based on off-resonant interaction
  followed by a measurement on light and a successive feedback onto
  the atoms.  This approach is particularly suitable for room
  temperature atomic ensembles because it can achieve a high fidelity
  for the so-called symmetric atomic spatial mode, that is the mode in
  which every atom contributes in the same way irrespectively of its
  position in the cell. For Raman and EIT based memories, a high
  fidelity can be achieved only for asymmetric atomic modes. The
  information is stored in a spin wave that is not distributed
  evenly over the ensemble so that atoms in the front part of the cell
  contribute much more to the memory than the atoms in the rear part
  of the cell \cite{novikova:243602}. Obviously such a memory can only
  have a life time that is limited by atomic motion, which is
  practically limited to a few tens of microseconds.

  In this project we use large ensembles of up to $10^{12}$ cesium
  atoms that are kept in glass cells which are coated from the inside
  with a material preventing collisional decoherence.  Our approach
  uses a quantum non-demolition (QND) type off-resonant interaction
  combined with quantum feedback. This has proven to work very well
  for the generation of deterministic entanglement of two distant
  objects \cite{julsgaard01:_entanglement}, for writing a coherent
  state of light into a quantum memory\cite{julsgaard04:_memory}, and
  for teleportation of a quantum state of light onto an atomic memory
  \cite{sherson06:_teleportation}.  Potential applications include
  distant teleportation of atomic states and high fidelity quantum
  memories for light. Since high fidelity teleportation is the
  essential ingredient of a quantum repeater this approach is also
  relevant for this application.

  \paragraph{Fidelity:} The maximum obtainable unconditional fidelity
  of the light-to-atoms storage increases with the resonant optical
  depth which for a detuning larger than Doppler width is equal to the
  optical depth for atoms at rest. An optical depth of the order of 50
  has been realized in experiments with Cesium atoms in a
  25 mm long cell at a temperature close to room temperature.
  Fidelities of storage and teleportation up to 70\% have been
  demonstrated experimentally for weak coherent states.  For atoms
  initially prepared in a squeezed state the fidelity can be even
  higher.  Such initial spin squeezing of the atoms can be achieved
  either by a quantum nondemolition measurement, or by preparing each
  atom in a coherent superposition of magnetic sublevels of the ground
  state.  With the latter method a spin squeezing of
  $\approx$
  3 dB has been demonstrated \cite{fernholz:073601}.

  An alternative approach to light-to-atoms state transfer is via
  light-to-atoms teleportation\cite{sherson06:_teleportation}. This
  approach has shown a fidelity of $\approx$ 60 \%, compared
  to the theoretical maximum of 72\%. Again, in order to
  improve the fidelity beyond this bound, initial squeezing should be
  employed, but this time the light which generates the entanglement
  should be squeezed.  With 6 dB of squeezing, which has been
  demonstrated experimentally in our laboratory, a fidelity exceeding
  90 \% is possible.

  The factors which reduce the experimentally observed fidelity below
  the theoretical maximum are losses of light and decoherence of
  atoms.  Losses of light are mainly caused by reflections on the cell
  windows. While on the external windows' surfaces dielectric
  antireflection coatings can be applied easily, the internal surfaces
  are difficult to coat.  However, it is clear that this purely
  technical limitation can be overcome. The atomic decoherence which
  is the most serious limitation to the fidelity is dominated by the
  decoherence caused by the driving optical field itself. At the
  moment this effect is responsible for the 10-15\% reduction in the
  fidelity compared to the theoretical optimum.

  \paragraph{Efficiency:} So far our most successful quantum memory
  implementations involved demonstration of unconditional state
  transfer from light onto atoms. The unconditional character of the
  memory is due to the fact that the measurement on light is a
  homodyne measurement on a well defined spatial mode of light which
  can be performed with detectors with more than $98\%$ quantum
  efficiency. The given fidelity is unconditional, and therefore by
  definition the efficiency is 100\%, i.e.  the process is
  deterministic.

  \paragraph{Storage time:} In our experiments the memory life time is
  defined as the time over which the fidelity drops below the
  classical benchmark value of 50\%.  Life times of up to 4 ms
  have been demonstrated experimentally \cite{julsgaard04:_memory}.
  Such a long quantum memory life time (the longest demonstrated so
  far for any kind of general quantum memory for light) has been made
  possible by using a special coating of the inside surface of the
  cells. With this coating atoms can withstand tens of thousands of
  collisions with the cell walls without loosing their coherence
  properties.

  \paragraph{Bandwidth:} The biggest drawback of our approach is that
  as of now only very narrowband light modes can be stored. The
  protocol relies on atoms to traverse the coupling beam many times
  during the write- and readout process, so that atom position effects
  average out.  If vapour cells of a few cm$^3$ are used this
  limits the bandwidth to $\approx$ 1 kHz. In order to
  construct a photon-qubit memory using this approach, ways to design
  a memory which can handle a bandwidth of about 1 MHz or
  higher should be developed. This could be achieved with smaller
  cells, or with hollow fibers.

  \paragraph{Multiple-photon and multiple-mode storage capacity:}
  The memory can store continuous variable observables with up to a
  few hundred photons and therefore the dimensionality of the stored
  state is high. In principle, several temporal modes could be stored
  in the same memory using different atomic sublevels. Limits are
  imposed by the number of levels available.

  \paragraph{Wavelength:} The wavelength should be close to resonant
  frequencies of alkali atoms, i.e., 852 nm or 895 nm
  for Cs and 780 nm or 795 nm for Rb, to take two most
  popular examples.  Those wavelengths are suitable for free space
  propagation. In order to apply this memory approach to telecom
  wavelengths for long distance fiber communications, the same
  approach can be, in principle, generalized to other systems or
  dual-color sources of entangled light states could be used.

  \paragraph{Read-out delay:} The limit of the read-out delay time is
  set by the quantum memory life time, which is up to 4 ms
  at the moment.

  \paragraph{Complexity:} The implementation uses room temperature
  atoms and therefore is simple and scalable.  As for all approaches
  that use a great number $N_a$ of individual physical systems to
  enhance the coupling between light and matter, if fields are used to
  perform local operations on states stored within the memory they
  have to be controlled precisely to $1/\sqrt{N}$. Very good magnetic
  shielding and low noise lasers and detectors are required to avoid
  technical noise.

\subsection{Cold trapped atomic ensembles (Copenhagen)}

  \paragraph{Brief description of approach and potential applications:}
  To some degree this approach is an extension of the room temperature
  experiments with alkali atoms: An optically dense cloud (optical
  depth of $\sim$15) of $10^5$ Cesium atoms is prepared in one of the
  magnetically insensitive clock-levels by first cooling the atoms in
  a magneto-optical trap and successively transferring them into a far
  off-resonant dipole trap, which is overlapped with one arm of a
  Mach-Zehnder interferometer\cite{windpassinger:103601}.
By measuring the the state- dependent phase shift that the
atoms induce on probe light in the interferometer  a QND
interaction between light and atoms can be realized with a
shot-noise and projection- noise limited precision
\cite{appel09:_spinsqueez}.

  The advantage of using a dense ensemble of cold atoms is that the
  sample is much smaller, light can be focused tighter and the same
  strong coupling can be achieved with much less photons and much
  shorter pulses compared to room temperature ensembles. Additionally
  the spontaneous scattering channels are closed, so that for similar
  interaction strengths QND measurements are much less destructive
  compared to the room-temperature vapor cell setup.

  Higher optical depth can be achieved with even colder or
  quantum degenerate atomic samples. Utilizing condensed
  samples containing $10^6$ Rubidium atoms stored in magnetic
  traps or optical dipole traps a considerable increase in
  coupling strength (optical depth of $\sim1000$), albeit at the
  cost of higher system complexity, is anticipated
  \cite{Hilliard:superradiance}. The most promising applications
  for cold atom based systems are in entanglement enhanced sensing
  and metrology and in quantum repeaters.

  \paragraph{Fidelity:} Fidelities similar to those already obtained
  with atomic vapors can be expected.

  \paragraph{Efficiency:} Like in the vapor-cell experiment the
  efficiency of the approaches we use is 100 \%, i.e. the
  process is deterministic.

  \paragraph{Storage time:}
  Quantum memory life times up to tens of milliseconds can be expected
  with ultracold atoms and $T_2$ times in this range are routinely
  observed for the collective coherence on the microwave clock transition. Recent
  investigations indicate that transversal motion of the atoms within
  the inhomogeneous probe beam \cite{oblak08:_echo_gauss} might reduce
  the life time of stored spin-waves, so that it might be necessary to
  employ an optical 3d-lattice instead of a simple linear dipole trap.

  \paragraph{Bandwidth:}
  Pulses of up to a few MHz bandwidth can be used. An upper
  limit is in principle given by the collective linewidth of
  the optical transition involved in the coupling scheme, so
  that for condensed samples also a bandwidth in the GHz range
  is possible.

  \paragraph{Multiple-photon and multiple-mode storage capacity:}
  Ensembles of 3d-trapped ultracold atoms are suitable for storage of
  multiple spatial modes.  In principle, several temporal modes could
  be stored in the same memory using different atomic sublevels.
  Limits are imposed by the number of levels available and the level
  of control/compensation of external fields that cause decoherence
  between the different hyperfine states. For spatial multimode
  operation ensembles with a sample Fresnel number above 1 need
  to be used. To prevent mode-mixing by atomic motion, strong
  localization in an optical lattice is advantageous.

  \paragraph{Wavelength:} The wavelength requirements are identical to
  those of room-temperature atomic ensembles and determined by
  the atomic species used.

  \paragraph{Read-out delay:} As in the room-temperature atomic vapors
  the limit of the read-out delay time is set by the quantum memory
  life time.

  \paragraph{Complexity:} The complexity is comparatively high: the
  implementation uses ultracold atoms trapped in a dipole trap and
  requires a very low phase noise microwave source to perform LOCC
  operations on the stored quantum state, as explained in
  Sec.~\ref{sec:room-temp-alkali}.

\subsection{Raman memory in atomic gases and solids
(Oxford)}

Off-resonant Raman interactions provide a natural way to
access long-lived material coherences optically, utilizing
the strong light-matter coupling of an intermediate dipole
transition, while avoiding fluorescent losses. Spontaneous
Raman scattering has already been used to entangle
separated atomic ensembles
\cite{Laurat:2007sw,Matsukevich:2005kr}, which operation
forms the basis of the DLCZ quantum repeater architecture
\cite{DLCZ}. Running the interaction in reverse offers the
possibility to coherently store propagating photons as
stationary excitations of a Raman coherence.

Our experiments have focussed on the manipulation of
temporally short, broadband wavepackets. We implemented a
Raman memory for sub-nanosecond pulses in cesium vapour,
and we probed the optical phonon modes of diamond via
ultrafast Stokes and anti-Stokes scattering. In both cases
coupling over a wide spectral bandwidth is possible because
the narrowband atomic states are dressed by the application
of a bright, short control pulse, producing a broad virtual
state that mediates the interaction.

\subsubsection{GHz optical memory in cesium vapour} 
\begin{figure}
\resizebox{\columnwidth}{!}{\includegraphics{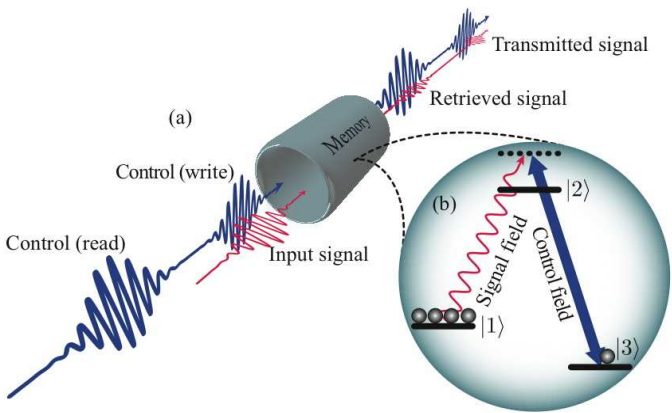}}
\caption{(a) A Raman memory. The signal
is directed into the memory along with a bright write pulse
and is stored. If the storage is partial, any unstored
signal is transmitted through the memory. A subsequent read
pulse extracts the stored excitation, which emerges along
with the transmitted read pulse. (b) The $\Lambda$-level
structure of the atoms in the memory. The atoms are
prepared in the ground state $|1 \rangle$ by optical
pumping. The signal is tuned into two-photon resonance with
the control field; both are detuned from the excited state
$|2 \rangle$. Absorption of a signal photon transfers an
atom from $|1 \rangle$ into the storage state $|3 \rangle$
via Raman scattering stimulated by the control. Upon
retrieval the interaction is reversed.}
\label{fig:Raman_1} 
\end{figure}

Although
Raman storage is well-known theoretically
\cite{Kozhekin,Nunn:2007wj,Gorshkov:2007th,Mishina:2008oq},
it has not previously been realized experimentally.
Superficially it resembles EIT-based storage, except that
the optical fields are detuned away from the excited state
resonance (see part (b) of figure \ref{fig:Raman_1}).
Qualitatively, they operate on quite different principles:
in EIT the strong dispersion associated with the quantum
interference of two absorption pathways --- direct or via
the control field --- is employed to bring a signal pulse
to a halt inside an atomic ensemble. In a Raman memory the
large detuning destroys this interference, and instead the
signal is coherently absorbed into the virtual state
created by the control pulse, with no reduction in group
velocity.

\begin{figure}
\resizebox{\columnwidth}{!}{\includegraphics{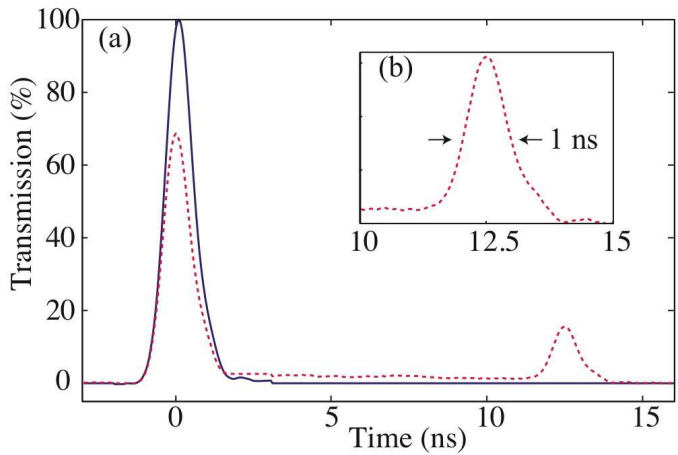}}
\caption{Storage and retrieval of a sub-nanosecond signal
pulse. (a) Solid line: Transmission of incident signal
field without the presence of control field --- there is no
storage and no retrieval. Dashed line: In the presence of
control fields (not visible) we observe 30\% storage and
50\% retrieval, yielding a total memory efficiency of
$15\%$. (b) Zoom of retrieved signal field showing the
measured full width at half maximum (FWHM) temporal
duration of 1 ns, limited by the detector response time.
This shows that the bandwidth of the retrieved signal
exceeds 1~GHz.} \label{fig:Raman_2} \end{figure}

In the experiment \cite{Reim2009}, a bright control pulse and an
orthogonally polarized weak signal pulse, both of $\sim
300$ ps duration, are spatially and temporally overlapped
and directed into a glass cell containing warm cesium
vapour, where around $30\%$ of the signal field is
converted into a \emph{spin wave} --- a stationary
excitation of the ground state hyperfine coherence. $12.5$
ns later, a second control pulse extracts the stored
excitation with $\sim 50\%$ efficiency, which emerges as a
retrieved signal pulse. Figure \ref{fig:Raman_2} shows a
typical time-trace of the intensity of the transmitted
signal field detected by a fast avalanche photodiode placed
downstream from the cell, after removal of the transmitted
control field by polarization and spectral filtering. It is
clear from part (b) that the bandwidth of the retrieved
signal field is at least $1$ GHz, but this measurement is
limited by the response time of the detector, which is $1$
ns. Theoretically the Raman memory operates at the full
bandwidth of the control pulse of around $1.5$ GHz.

Decoherence of the spin wave over the short storage time shown here is negligible; recently we observed retrieval after $\sim 2 \mu$s, although some technical adjustments are required to test longer storage times. The efficiency of our memory is limited by
the strength of the Raman coupling attainable, and by the
quality of the overlap between the write control pulse and
the incident signal. Appropriate pulse shaping (which may
be technically challenging), and increasing the control
pulse energy (which should not be), could raise the total
efficiency to around $60\%$ \cite{Nunn:2007wj}.
Phasematched retrieval in the backward direction allows
efficiencies above $90\%$ to be reached
\cite{Surmacz:2008vf}.

The memory performs reasonably well when evaluated against
the criteria introduced in Section \ref{criteria}. The {\bf
fidelity} of the memory is (up to a possible unitary
re-shaping of the retrieved pulse profile
\cite{Surmacz:2006rz}) the same as the {\bf efficiency},
which is $\sim 15\%$. As discussed above, much higher efficiencies are feasible. A proxy for the {\bf conditional fidelity} is given by the visibility of interference between the retrieved pulse and an appropriately attenuated replica of the stored pulse. The short storage time of 12.5 ns makes this measurement relatively straightforward, and we obtained a visibility of $\sim 85\%$, after correcting for imperfections in the interferometer. Theory suggests further improvement may be possible if the Stark shift on the retrieved signal due to the strong control can be compensated.

The limiting {\bf storage time} of the current memory is probably on the order of 100 $\mu$s; a typical coherence time in warm atomic vapours set by the diffusion of atoms out of the interaction region \cite{Camacho:2009ao}. Longer times are of course possible if the atoms can be confined, for example by the walls of the vapour cell, although special care should be taken that the optical fields address the full cell diameter, and that collisions of atoms with the cell wall do not dephase the Raman coherence. Coherence times in warm cesium vapour on the order of 10 ms have been achieved with such a set-up by the Copenhagen group \cite{julsgaard01:_entanglement}, with only residual magnetic fields and Cs-Cs spin exchange preventing longer coherence times. Other possibilities for extending the memory lifetime include trapping the atoms in an optical lattice \cite{Schnorrberger} or — most appealing of all — as dopants in a solid-state host (see the discussion of CRIB and AFC memories in section 6.1).

The {\bf bandwidth} of the current memory exceeds $1$ GHz,
which compares favourably with the clock rates of modern
classical computers. Time bandwidth products of the order
of $10^5$ are feasible if the full storage time is
exploited. The bandwidth is limited by the ground state
hyperfine splitting of $9.2$ GHz in cesium, since larger
bandwidth pulses will address both control and signal
transitions simultaneously (this is known as the impulsive
regime in the Raman literature \cite{Bloembergen:1996la}).
Other media with larger splittings, or appropriate
selection rules, could allow the storage of shorter pulses;
the diamond memory introduced in the next section is an
example.

The {\bf multimode capacity} is not generally large --- it
scales poorly with the density of the ensemble, as indeed
does the capacity of EIT storage \cite{Nunn:2008oq} ---
although off-axis beam geometries do allow for angular
multiplexing
\cite{Surmacz:2008vf,Tordrup:2008rm,Vasilyev:2008wm}.
Finally, the {\bf wavelength} is tunable in principle,
although the present memory is operated sufficiently close
to resonance that significant wavelength variations away
from $852$ nm would rapidly erode the efficiency: the
tunability is currently limited to a few GHz. Again,
alternative media with sufficiently large Raman
cross-sections --- such as the diamond memory discussed
below --- can be operated far from resonance, where the
Raman coupling varies slowly with frequency, allowing for
near-arbitrary wavelength tunability.

\paragraph{Solid state memory in diamond}
Diamond has a very large Raman cross-section. Although its
bandgap lies in the ultraviolet, optical pulses will
readily scatter from the `deformation potential', producing
correlated pairs of Stokes photons and optical phonons.
These phonons are non-propagating vibrational excitations
associated with the `ringing' of the crystal basis. They
survive for approximately $5$ ps before decaying into pairs
of short wavelength acoustic phonons --- sound waves ---
due to high order anharmonic couplings in the diamond
lattice. Although this lifetime is extremely short, the
optical phonons are interesting because they are very
energetic, with a Stokes shift of $1332$ cm$^{-1}$ ($0.16$
eV, or expressed as an optical wavelength, $7.5$ $\mu$m).
This is much larger than the energy scale associated with
thermal fluctuations at room temperature, so these
excitations are naturally isolated from noise.
Additionally, the large splitting allows Stokes photons to
be spectrally distinguished from the control field used to
initiate the interaction, even if their spectra span $10$s
of nanometers. Therefore, ultrashort, femtosecond timescale
pulses can be used to address the phonons: on such
timescales the coherence time of the phonons appears long!

\begin{figure}
\resizebox{\columnwidth}{!}{\includegraphics{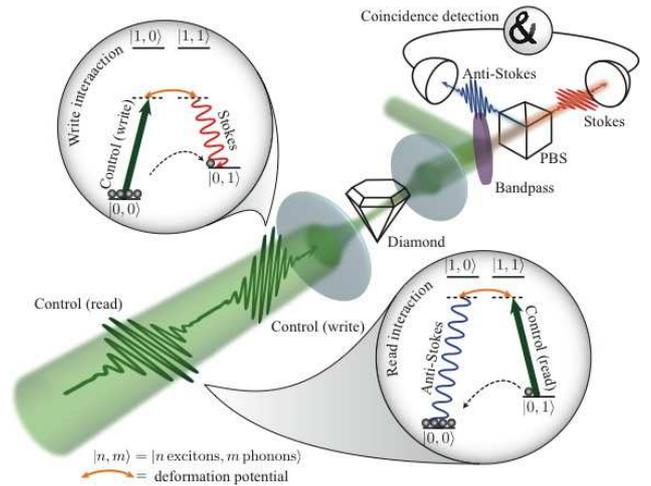}}
\caption{Creation and detection of phonon excitations in
diamond. A pair of $50$ fs control pulses are focussed into
a crystal of diamond. When the `write' pulse generates a
Stokes photon, a phonon is created. The `read' pulse can
then scatter from this excitation, producing an anti-Stokes
photon that returns the crystal to its ground state.
Coincidence measurements reveal the resulting statistical
correlations arising between Stokes and anti-Stokes
detection events.} \label{fig:diamond_T1} \end{figure}

The optical phonons in bulk diamond represent a unique
testbed for the ultrafast control of macroscopic coherence
in solid state systems at room temperature. As an extension
of previous work \cite{kuzmich,KimbleNature} in atomic
vapours to such systems, we adopted the goal of generating,
and then verifying, entanglement between the phonon modes
of two separated diamond crystals. We have successfully
taken the first step in this direction, namely
demonstration of the ability to create and detect phonons.

\begin{figure}
\resizebox{\columnwidth}{!}{\includegraphics{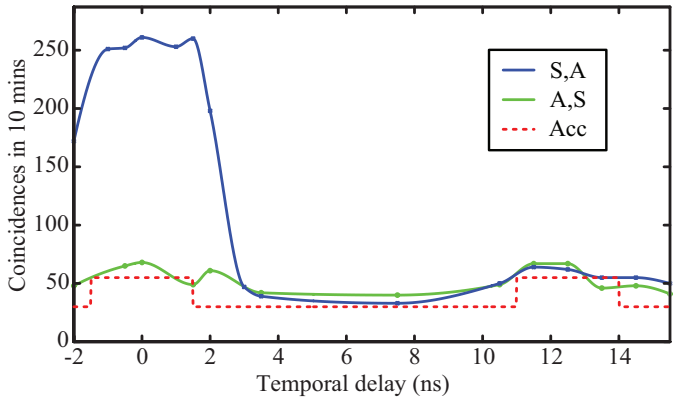}}
\caption{Read out of phonons in diamond, as revealed by
examining the correlation between Stokes and anti-Stokes
detection events. The green curve shows the coincidence
rate associated with the `wrong' detection order (A,S), in
which an anti-Stokes photon is scattered from the write
pulse and a Stokes photon scatters from the read pulse.
Photons scattered from the write and read pulses are
distinguished by their polarizations. The blue curve
describes the correlations for the `correct' ordering
(S,A). No significant correlations are observed for the
incorrect ordering (although the polarization selection
rules in diamond are not perfect, so there is a residual
contribution from the correct ordering). The correct
ordering exhibits a pronounced peak around zero delay. This
peak is a signature that phonons are created by the write
pulse and read out by the read pulse from the same control
pulse sequence --- the  $5$ ps delay between the read and
write control pulses is not resolvable. Subsidiary maxima
occur for delays that are multiples of $12.5$ ns, which is
the laser repetition rate. These are spurious, being a
symptom of the increased accidental coincidence rates
associated with a pair of independent control pulse
sequences (dashed red line, calculated for a $3$ ns
coincidence gate, with reasonable technical noise).}
\label{fig:diamond_2} \end{figure}

Figure \ref{fig:diamond_T1} shows the experimental
arrangement used, along with the structure of the
interactions involved. In figure \ref{fig:diamond_2} the
Stokes/anti-Stokes coincidence rate is plotted as a
function of an electronically introduced time-delay between
the two detection events. The surge in correlations around
zero delay is attributed to the survival of phonons over
the (unresolved) period separating the control pulses.

The next stage of these experiments is to interfere the
Stokes modes from a pair of diamond crystals, so as to
erase the \emph{welcher weg} information about the origin
of a detected Stokes photon. Such a detection then
entangles the phonon modes of the crystals; subsequent
application of read pulses to the crystals transfers this
entanglement back to the optical domain, where it can be
witnessed by interfering the anti-Stokes modes. The scheme
is identical to that implemented in
\cite{Chaneliere:2007pb,kuzmich,Laurat:2007sw}, except that
the entanglement is created in a solid, a technical
challenge being that is survives for a time that is too
short to be resolved electronically.

Although the current experiments with diamond are not
focussed on the realization of an absorptive optical
memory, the planned entanglement generation set-up can be
adapted to function as a post-selected memory
\cite{kuzmich}, which is prepared by the Stokes detection
that heralds entanglement. From this perspective, it is
interesting to evaluate the performance of diamond as a
memory by the criteria introduced in Section
\ref{criteria}.

For storage of a polarization qubit, the {\bf fidelity} can
be high in principle, being limited only by the stability
of the set-up and the coherence time of the memory. The
retrieval probability is generally low, which translates
into low {\bf efficiency}, although this could be mitigated
by boosting the energy of the read pulse. The {\bf storage
time} of $5$ ps is extremely short --- far too short to
find applications in any quantum communication protocols.
It is possible, however, that it could be of use as a
solid-state register, as part of a miniaturized integrated
quantum processor.

The {\bf bandwidth} is extremely large, being on the order
of tens of THz. This allows the storage of ultrafast
pulses, a million times shorter than any other memory has
so far demonstrated. The time bandwidth product for the
memory is accordingly on the order of $1000$, despite its
short coherence lifetime.

The {\bf multimode capacity} of the memory is probably
small, for the same reasons as discussed above in the case
of an absorptive Raman memory. Nonetheless, as with all
ensemble memories, spatial or angular multiplexing could be
used. Finally the {\bf wavelength} is extremely tunable,
since diamond is transparent to all optical frequencies,
with a Raman cross-section that varies little over the
visible or telecoms spectra. Essentially any choice of
wavelength is viable.

\section{Summary and Outlook}
\label{summary}

\begin{figure*}[h!]
\resizebox{1.67\columnwidth}{!}{\includegraphics{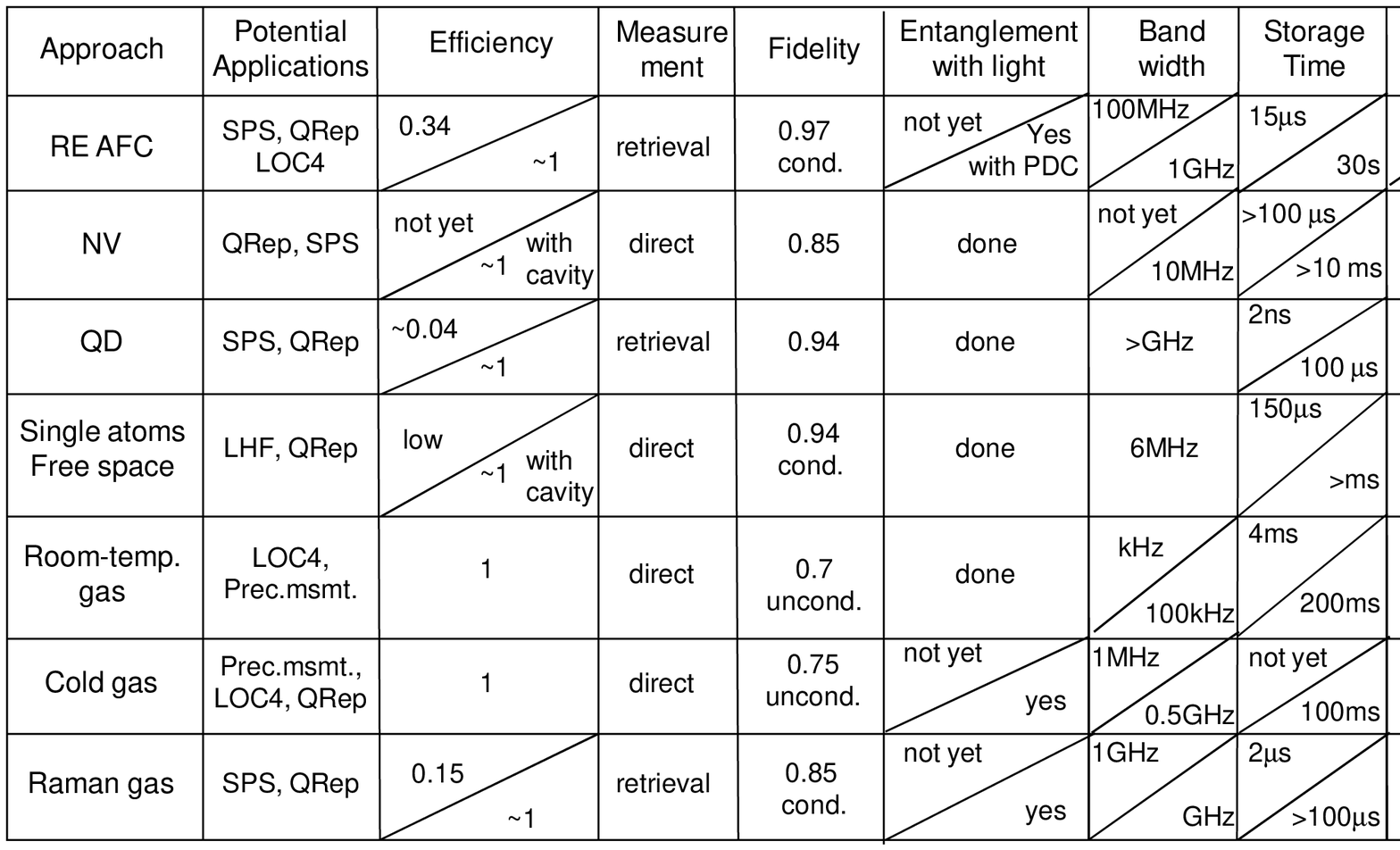}}
 \caption{Overview of different approaches to quantum memories represented in the Integrated Projet {\it Qubit Applications}. The numbers shown are the best values that have been achieved by members of the project. {\bf Approach:} RE AFC - Atomic Frequency Comb with Rare-Earth Doped Crystals (Geneva/Lund); NV - NV centers in diamond (Stuttgart); QD - Quantum Dots (Toshiba); Single atoms in free space (Munich); Room-temperature gas
(Copenhagen); Cold gas (Copenhagen); Raman gas - Raman
memory in a hot Cesium gas (Oxford). {\bf Potential
Applications:} See section 2 for a discussion of potential applications. SPS - Single-Photon Source; QRep - Quantum Repeaters; LOC4 - Protocols that allow a reduction in
Communication Complexity using Memories and Local
Operations and Classical Communication; LHF - Loophole-Free
Bell experiment; Prec.msmt. - Precision measurement. {\bf
Efficiency:} See sections 4.2 and 6 for the concept of efficiency. Above the diagonal is the
current experimental value, below the diagonal the value
that appears in principle achievable with the respective
approach. The same applies to the following columns. {\bf
Measurement:} In some approaches it is possible
to measure the stored state in a way that is different from
retrieving the stored light. In other approaches state
measurements are done via retrieval. {\bf Fidelity:} See sections 4.1 and 6 for the definition of fidelity. The given fidelities are conditional on detection of a photon, or unconditional as
indicated.  {\bf Entanglement with Light:}
Entanglement between an excitation stored in the memory and
light has already been demonstrated in some approaches.
{\bf Bandwidth:} Bandwidth of the light that can be stored
and retrieved (or that is emitted, cf. section 3.3). {\bf Storage Time:} Maximal memory storage
time. {\bf MM capacity:} Capacity to store signals in
multiple modes. {\bf Dim.:} Capacity to store signals that
live in a high-dimensional Hilbert space (for a single mode).}
\label{table}
\end{figure*}

Fig. \ref{table} summarizes where the different approaches
position themselves with respect to the applications
discussed in section 2 and to the criteria discussed in
section 4.

The experimental projects in this review are in different
stages in their development. Some have already demonstrated
quantum features such as entanglement with light or
realized quantum protocols such as teleportation. Others are in an earlier stage. All approaches appear to be suitable in principle for high-fidelity and high-efficiency operation. The technological challenges
that have to be overcome to this end vary between the
different approaches. Most implementations aim for
bandwidths in the range 100 MHz to 1 GHz and for storage
times in the ms to s range. Concerning complexity, most
solid-state implementations require cryostats and some
micro- or nanofabrication, while atomic approaches tend to
require laser cooling and trapping.

In the future it will be interesting to investigate various
ways of combining the different approaches discussed here.
Recent examples for this trend include the combination of
continuous-variable and single-photon concepts in the
context of quantum repeaters \cite{Sangouardcohstates}, and
a proposal for creating entanglement between a trapped ion
and a quantum dot \cite{waks}.

{\bf Acknowledgements.} The work on this review paper was supported by the EU Integrated Project {\it Qubit Applications}. We are grateful to G. Alber, T. Chaneli\`{e}re, K. Hammerer, N. Korolkova, J.-L. Le Gou\"{e}t, A. Lvovsky, K. M{\o}lmer, B. Sanders, N. Sangouard, A. S{\o}rensen, K.-A. Suominen, and W. Tittel for useful discussions and helpful comments.

%
%
%
%

\end{document}